%% file: rrrNetworksMain-arxiv.tex
\documentclass[10pt, one column, draft cls]{ieeeconf}
\IEEEoverridecommandlockouts 

\input{config-arxiv}

\input{./inputsJSS/configJSS}

\graphicspath{{../../../../fig/}{../fig/fig/}{../fig/tikz/}}

\title{Network Effects on Robustness of Dynamic Systems}

\author{Ketan Savla\thanks{
Ketan Savla is with the University of Southern California, Los Angeles, CA, USA, \texttt{ksavla@usc.edu}.
Jeff Shamma is with the King Abdullah University of Science and Technology, Thuwal, Saudi Arabia. \texttt{jeff.shamma@kaust.edu.sa}.
Munther Dahleh is with the Massachusetts Institute of Technology, Cambridge, MA. \texttt{dahleh@mit.edu}.}, Jeff S. Shamma, and Munther A. Dahleh}

\date{}							

\begin{document}
\maketitle

\input{abstract}

\input{introduction-revised}


\input{notations}
\input{./inputsJSS/notationsJSS}

\input{./capacity-static/dc-networks-v2}


\input{./non-asymptotic-dynamic/non-asymptotic-dynamic-main-v2}


\input{./capacity-dynamic/network-flow-dynamics-v2}


\input{./cascading-failure/cascading-failure-main-v2}


\input{./asymptotic-static/asymptotic-static-main}


\input{./asymptotic-dynamic/consensus}

\input{conclusion}

\section*{ACKNOWLEDGMENTS}
This work was supported by NSF CAREER ECCS \# 1454729, and by funding from King Abdullah University of Science and Technology (KAUST). The authors thank Bassam Bamieh for helpful discussions. 

\bibliographystyle{IEEEtran}
\bibliography{./network-robustness,./inputsJSS/refsJSS}

\end{document}

%% file: config-arxiv.tex
\usepackage{amsmath,amssymb,bbm,color,psfrag,amsfonts,bm,amsbsy}
\usepackage{graphicx}
\usepackage{pdfpages}
\usepackage{standalone}
\usepackage{tikz}

\input{./tikz_config}

\newtheorem{theorem}{Theorem}[section]

\newtheorem{proposition}{Proposition}[section]

\newcommand{\mc}{\mathcal}

\newcommand{\ksmarginsoft}[1]{}

\newcommand{\bigtriangle}{\pmb{\triangle}}

\newcommand{\kscommentjss}[1]{#1}
\newcommand{\zerobf}{\mathbf{0}}
\newcommand{\onebf}{\mathbf{1}}

\newcommand{\kscommentsoft}[1]{#1}
\newcommand{\reals}{{\mbox{\bf R}}}
\newcommand{\naturals}{{\mbox{\bf N}}}
\newcommand{\E}{\mathbb{E}}
\newcommand{\var}{\mathrm{var}}
\newcommand{\map}[3]{#1: #2 \rightarrow #3}

\newcommand{\Azero}{A^{(0)}}

\newcommand{\Hop}{\mathbb{H}}

\newcommand{\linkin}{{\mc E}^{\text{in}}}
\newcommand{\linkout}{{\mc E}^{\text{out}}}
\newcommand{\interflow}{z}
\newcommand{\interflowconstraints}{\mc Z}
\newcommand{\incidence}{B}
\newcommand{\diag}{\mathrm{diag}}
\newcommand{\biglambda}{\Lambda}
\newcommand{\lambdamax}{\lambda^{\max}}
\newcommand{\Wmin}{\underline{\mc W}}
\newcommand{\Wmax}{\overline{\mc W}}
\newcommand{\Weq}{\mc W^{\mathrm{eq}}}
\newcommand{\argmax}{\mathrm{argmax}}
\newcommand{\failprob}{r}
\newcommand{\graphclass}{\mathbb{G}}
\newcommand{\triangletree}{\bigtriangledown}
\newcommand{\complex}{\mathbf{C}}
\newcommand{\shock}{s}
\newcommand{\energyavg}{\mathbf{E}}
\newcommand{\energymax}{\mathbf{M}}

%% file: tikz_config.tex
\usepackage{tikz}
\usepackage{pgfplots}
\usetikzlibrary{positioning}
\usetikzlibrary{shapes, arrows.meta, arrows}
\usepackage{relsize}

\tikzset{fontscale/.style = {font=\relsize{#1}}
    }

\definecolor{nodecolor}{rgb}{255, 255, 255}
\definecolor{sourcenodecolor}{rgb}{0, 0, 0} 
\definecolor{sinknodecolor}{rgb}{0, 0, 0} 

\tikzset{ dot/.style ={circle, draw, inner sep=2pt},  
	     supply dot/.style ={circle, draw, inner sep=2pt, fill = sourcenodecolor}, 
	     demand dot/.style={circle, draw, inner sep=2pt, fill = sinknodecolor},
              main node/.style={circle,draw,font=\sffamily\bfseries\small, fill=nodecolor, line width = 0.5mm}, 
              supply node/.style={circle,draw,font=\sffamily\bfseries\small, fill=sourcenodecolor, line width = 0.5mm}, 
              demand node/.style={circle,draw,font=\sffamily\bfseries\small, fill=sinknodecolor, line width = 0.5mm},
              edge label/.style={font=\sffamily\small},
              main edge/.style={thick, auto},
              directed edge/.style={-Stealth,thick, auto}, 
              cascading edge/.style={line width=0.75mm, auto},
              infeasible node/.style={circle,minimum size=0.6cm, inner sep=0pt, draw, fill= red!50}, 
              feasible node/.style={circle,minimum size=0.6cm, inner sep=0pt, draw, fill= blue!50},
              face/.style={circle,draw,inner sep=1pt, fill=black}}


%% file: inputsJSS/configJSS.tex
\newcommand{\theset}[1]{\left\{ #1\right\}}
\newcommand{\Ltwo}{\mathbf{L}_2}
\newcommand{\LtwoE}{\mathbf{L}_{2,e}}
\newcommand{\arr}{\mathbf{R}}
\newcommand{\st}{\ \Big\vert \ }
\newcommand{\tr}{^\mathrm{T}}
\newcommand{\norm}[1]{\left\Vert #1 \right\Vert}
\newcommand{\magn}[1]{\left\vert #1 \right\vert}
\newcommand{\dt}{\thinspace \mathrm{dt}}
\newcommand{\forAll}{\text{ for all }}
\newcommand{\Pmatrix}[1]{\begin{pmatrix}#1\end{pmatrix}}
\newcommand{\sysMat}[4]{\left(\begin{array}{c | c} {#1} & {#2}\\ \hline {#3} & {#4} \end{array}\right)}

%% file: abstract.tex
\section*{Abstract}

We review selected results related to robustness of networked systems in finite and asymptotically large size regimes, under static and dynamical settings. 
In the static setting, within the framework of flow over finite networks, we discuss the effect of physical constraints on robustness to loss in link capacities. In the dynamical setting, we review several settings in which small gain type analysis provides tight robustness guarantees for linear dynamics over finite networks towards worst-case and stochastic disturbances. We also discuss network flow dynamic settings where nonlinear techniques facilitate in understanding the effect on robustness of constraints on capacity and information, substituting information with control action, and cascading failure. We also contrast the latter with a representative contagion model. For asymptotically large networks, we discuss the role of network properties in connecting microscopic shocks to emergent macroscopic fluctuations under linear dynamics as well as for economic networks at equilibrium. Through the review of these results, the paper aims to achieve two objectives. First, to highlight selected settings in which the role of interconnectivity structure of a network on its robustness is well-understood. Second, to highlight a few additional settings in which  existing system theoretic tools give tight robustness guarantees, and which are also appropriate avenues for future network-theoretic investigations.

%% file: introduction-revised.tex
\section{Introduction and Overview}
Robustness is the ability of a system to operate effectively under a range of different environmental conditions
or in the face of possible disruptions. What is the impact of incidents on traffic flow? How does an energy grid respond to surges in demand?
How do supply chain disruptions impact a production economy? Under what conditions can a 
communication network maintain quality of service?  Of course, the issue of robustness is at the heart of control systems and a primary motivation for the introduction of feedback in dynamical systems.

While the concept of robustness is widely relevant to multiple domains, an application area
of particular importance is networked systems, including physical networks; financial, economic and social networks; or networks of people and systems.
Indeed, tremendous advances in communication, computation, and sensing have led to renewed interest in networked
systems, in that new technologies are introducing unprecedented interdependencies between people, devices, and infrastructure.

In such settings where the operation of one component impacts the operation of other components,
the presence of a network structure significantly impacts the characterization of robustness. 
While each component of a networked system can experience its own disruptions, the impact on the overall 
system depends critically on the specific structure of the interconnectivity. Stated differently, the impact of the same component-level disruptions
may range from inconsequential to critical, depending on the network structure. This paper presents a review of selected results on this issue of network effects on robustness. 

Section~\ref{sec:capacity-perturbation-static} begins the discussion by considering the specific setting of physical flow over networks. There is a network of links, each with limited capacity, that must accommodate
the overall flow. Disruptions take the form of reductions in capacity. The focus is on a static problem of whether
a network has sufficient excess capacity to withstand the impact of an adversarial environment with a limited budget of capacity
reduction, and how to compute this excess capacity in an efficient manner. 

Section~\ref{sec:non-asymptotic-dynamical} presents analysis of robustness of feedback interconnections. The question is under what conditions does a feedback system maintain stability (and more generally, performance) in the presence of a
specified family of possible model perturbations. The framework is primarily motivated by modeling of physical systems, where model simplifications
are introduced for the sake of control design and analysis. The underlying network effect is reflected in the specifics of where model perturbations occur
in a feedback interconnection. The discussion is separated based on whether the 
%
model perturbations are deterministic (i.e., a worst-case analysis) and with memory, or they are stochastic and memoryless. A motivating application for the latter is a communication network with unreliable links, and the main results there express tight conditions under which the resulting feedback system is stochastically stable. 

Section~\ref{sec:capacity-perturbation-dynamic} examines the nonlinear dynamic setting in the specific context of network flow dynamics under capacity constraint. Network flow is controlled in a distributed way,
i.e., based on local conditions as opposed to centralized control with full information of network conditions. Specific settings are presented to illustrate the impact of such limited information on network robustness to loss in capacity, and how to compensate for information constraint with additional control action.  

Section~\ref{sec:cascading-failure} discusses robustness under cascading failure, under which the collapse of
one component in a network propagates to other components, again as determined by an underlying network architecture. Here the motivation is to model
the mechanism of failure and quantify network robustness.  The specific settings include electrical networks, transport networks, and contagion models.

Section~\ref{sec:asymptotic-static} is motivated by the question of how random disruptions impact a network. A general perspective is that the effects of multiple sources
of random disruptions average out to have a diminishing effect as the network grows. Alternatively, 
it may be that the network structure results in an amplification of such disruptions. These issues motivated the notion of
systemic risk in financial networks. In the context of an interconnected production economy, Section~\ref{sec:asymptotic-static} presents results that demonstrate that 
random component-level disruptions can indeed be amplified because of a network structure. This question is addressed in 
two ways, first in the size (variance) of the overall effect, and second, in terms of the probability of an extreme event (as captured by a specific notion of risk).
These results are ``static'' in that they apply to the equilibrium behavior of a dynamical system,
and  ``asymptotic'' in the sense that they capture the effects for progressively larger networks.

Finally, Section~\ref{sec:asymptotic-dynamical} also addresses the impact of disturbances, but focuses on the transient behavior (i.e., not just equilibrium) of a 
dynamical system. The specific setting is the energy of the response of a linear system to a disruptive initial impulse. Again the conclusions 
are asymptotic in terms of progressively larger systems (i.e., more states). The underlying network is captured within the
specific structure of the dynamics.

These results illustrate multiple approaches that one can take in analyzing robustness of networked systems,
e.g., static versus dynamic models, deterministic versus stochastic disruptions, or asymptotically large versus fixed size networks.
Nonetheless, the common theme throughout is understanding the network effect of how conclusions depend on the underlying
interdependency structure. Future research directions along these lines are also suggested at the end of each of above sections.

%% file: notations.tex
\section{Notations and Preliminaries}
\label{sec:notations}

\textbf{Miscellaneous}: The sets of real, non-negative real, and positive real numbers are denoted as $\reals$, $\reals_{\geq 0}$ and $\reals_{> 0}$ respectively. $\naturals$ denotes the set of natural numbers. 
For a set $S$, $|S|$ denotes its cardinality. $\reals^{S}$, $\reals_{\geq 0}^{S}$ and $\reals_{>0}^{S}$ will, respectively, be shorthand notations for $\reals^{|S|}$, $\reals_{\geq 0}^{|S|}$ and $\reals_{>0}^{|S|}$. 
The set of complex numbers is denoted as $\complex$. For a vector $v \in \complex$, $v^{H}$ denotes the complex conjugate transpose. 
$[m]$ is short hand for $\{1, 2, \ldots, m\}$. For functions $f(.)$ and $g(.)$, we have $f(n)=O(g(n))$, when there exist constants $C$, $n_0$ such that $f(n) \leq C g(n)$ for all $n \in \naturals > n_0$. If $f(n)=O(g(n))$, then $g(n)=\Omega(f(n))$. 
$f(n)=\Theta(g(n))$ when there exist constants $C_1$, $C_2$, $n_1$ such that $C_1 g(n) \leq f(n) \leq C_2 g(n)$ for all $n \in \naturals > n_1$.

\textbf{Probability Theory}:
$\E[X]$ and $\var[X]$ will respectively denote the expected value and variance of random variable $X$. 
$\Phi(.)$ will denote the cumulative distribution function of Gaussian distribution:
$$
\Phi(x) = \frac{1}{\sqrt{2 \pi}} \int_{- \infty}^{x} e^{-t^2/2} \, dt
$$
A random variable $X$ is said to exhibit \emph{tail risk} (relative to the normal distribution) if $\lim_{\tau \to \infty} r_X(\tau)=0$, where the \emph{$\tau$-tail ratio} of $X$:
\begin{equation}
\label{eq:tail-ratio}
r_X(\tau) = \frac{\log \Pr(X < \E[X]-\tau \sigma_X)}{\log \Phi(-\tau)}
\end{equation}
is the probability that $X$ deviates by at least $\tau$ standard deviations from its mean relative to similar probability of deviation under the standard normal distribution:
We say that $X$ is \emph{light-tailed} if $\E[\exp(bX)]<\infty$ for some $b>0$. Otherwise, we say $X$ is \emph{heavy-tailed}. Any heavy-tailed random variable exhibits tail risk, but not all random variables with tail risks are heavy-tailed. 

\ksmarginsoft{Is $\sigma$ consistent with $M-\triangle$ section?}
\textbf{Matrix Theory}: 
For a matrix $A$, $[A]_{ij}$ denotes its $(i,j)$-th element, $A_i$ denotes its $i$-th row, and $A(i)$ denotes its $i$-th column, and $\rho(A)$ denotes its spectral radius. $\sigma(A)$ will denote the singular value of $A$, which is greater than or equal to other singular values of $A$. A matrix is called non-negative, if all of its entries are non-negative. A non-negative matrix $A$ is said to have a \emph{Perron} root $\lambda_{\text{PF}}$ if $\lambda_{\text{PF}}$ is a positive real number such that it is an eigenvalue of $A$ and every other eigenvalue $\lambda$ of $A$ satisfies $|\lambda| < \lambda_{\text{PF}}$.
Given compatible matrices $A_1$ and $A_2$, $A_1 \circ A_2$ will denote their Hadamard, i.e., element-by-element, product.
$I_n$ is the $n \times n$ identity matrix. $\zerobf_n$ and $\onebf_n$ will, respectively, denote vectors of all zeros and ones of size $n$. We shall drop the subscript on size when clear from the context. Given two vectors $a$ and $b$ of the same size, $a \leq b$ would imply entry-wise inequality. $e_i$ will denote the column vector whose $i$-th entry is one, and other entries are zero; its size will be clear from the context.

\ksmarginsoft{should stick to either 'edge' or 'link' throughout the paper}
\textbf{Graph Theory}: A graph is the tuple $\mc G = (\mc V, \mc E, \mc W)$, where $\mc V=\{v_1, \ldots, v_n\}$ is the set of $n$ nodes, $\mc E$ is the set of edges, and $\map{\mc W}{\mc E}{\reals}$ assigns weights to edges. 
A directed edge from node $i$ to node $j$ is denoted by $(v_i,v_j) \in \mc E$; $v_i$ is referred to as the tail node, and $v_j$ the head node. $\mc E_v^+$ (resp., $\mc E_v^-$) will denote the set of edges outgoing from (resp., incoming to) node $v$, i.e., all the edges whose tail node (resp., head node) is $v$. The node-link incidence matrix $\incidence \in \{\pm1,0\}^{\mc V \times \mc E}$ is defined such that $\incidence_{ve}$ is equal to $-1$ (resp., $+1$) if $v$ is the head (resp., tail) node of $e$, and equal to zero otherwise.  An edge $(v_i,v_j)$ is said to be incident on to edge $(v_k,v_{\ell})$ if $v_j=v_k$. 
$\mc G$ is called symmetric or undirected if $\mc W(v_i,v_j) \equiv w_{ij} = w_{ji} \equiv \mc W(v_j,v_i)$ for all $1 \leq i, j \leq n$. A directed path from $v_i$ to $v_j$ is an ordered sequence of vertices $v_i, v_k, \ldots, v_j$ such that any pair of consecutive vertices in the sequence is a directed edge. $\mc G$ is said to be strongly connected if there exists a directed path from $v_i \in \mc V$ to $v_j \in \mc V$ for all $i, j \in [n]$, $i \neq j$, and is said to be weakly connected if the undirected version of $\mc G$ is strongly connected. $\mc G = (\mc V, \mc E, \mc W)$ is said to be induced by a matrix $A \in \reals^{n \times n}$ if $|\mc V|=n$, $(i,j) \in \mc E$ if $[A]_{ij} \neq 0$, and $\mc W(v_i,v_j)=[A]_{ij}$.\footnote{\kscommentsoft{We shall use $\mc W$ to denote both the map as well as the matrix whose entries are $w_{ij}$. When $\mc W$ is induced by $A$, we shall use $\mc W$ and $A$ interchangeably.}} \ksmarginsoft{better to replace matrix $\mc W$ with $A$ everywhere in the paper}
Conversely, given a graph $\mc G$, there exists a matrix $A$ which induces $\mc G$. Therefore, we shall use graph $\mc G$ and the associated matrix $A$ interchangeably to refer to the same object. Also, with a slight abuse of terminology, we shall use the terms \emph{graph} and \emph{network} interchangeably. 
Similarly, we shall use \emph{edge} and \emph{link} interchangeably. 
The degree of a node $v$ in an undirected graph is the sum of the weights of the links at $v$. If the graph is unweighted, then the degree of a node is simply the number of edges at $v$. If all the nodes of an undirected graph have the same degree $d$, then the graph is called $d$-\emph{regular}. An $(n-1)$-regular graph with $n$ nodes is called \emph{complete}. 
On the other hand, we distinguish between \emph{in}- and \emph{out}-degrees for a directed graph. The outdegree of a node is defined to be equal to the sum of the weights of the edges outgoing from that node; indegree is defined similarly. 

At times, we shall refer to \emph{large networks}, by which we shall mean a sequence of networks $\{\mc G_n = (\mc V_n, \mc E_n, \mc W_n)\}_{n \in \naturals}$ with matrices $\{A_n\}_{n\in \naturals}$, where the topology of each network in the sequence is fixed but the network dimension (i.e., number of nodes) grows successively. 

\textbf{Network Flow}: Consider a directed graph $\mc G = (\mc V, \mc E, \mc W)$, with specific two nodes $s, t \in \mc V$ designated to be \emph{source} and \emph{sink} respectively. Each link is associated with a \emph{flow} variable; let $\{f_{ij} \geq 0\}_{(i,j) \in \mc E}$ be the vector of link-wise flows. 
We associate each link $(i,j)$ with a \emph{flow capacity} $c_{ij}>0$, i.e., the flow variables are constrained to be $f_{ij} \leq c_{ij}$ for all $(i,j) \in \mc E$. The $c_{ij}'s$ are to be distinguished from link weights $w_{ij}'s$. 
Additionally, $f$ is constrained to satisfy flow conservation at every node $i \in \mc V \setminus \{s,t\}$: 
\begin{equation}
\label{eq:flow-conservation}
\sum_{j: \, (j,i) \in \mc E} f_{ji} = \sum_{j: \, (i,j) \in \mc E} f_{ij}
\end{equation}
The value of flow $f$ is defined to be equal to the difference between outflow and inflow at the source node, i.e., $\sum_{j: \, (s,j) \in \mc E} f_{sj} - \sum_{j: \, (j,s) \in \mc E} f_{js}$. The \emph{maximum flow problem} for a given $\mc G$, $s$ and $t$ is to maximize the value of flows over all $f$ satisfying the capacity and flow conservation constraints. This problem can be formulated as a linear program, e.g., see \cite[Chapter 8]{Korte.Vygen:02}. Given its importance for several applications involving large networks, developing computationally efficient algorithms for this problem and its variants continue to attract attention, e.g., see \cite{Christiano.Kelner.ea:ACM11}.  

Interestingly, the solution to the above maximum flow problem is equal to the solution of a combinatorial optimization problem, the \emph{minimum $s-t$ cut problem}, over the same data. The solution to the minimum $s-t$ cut problem is equal to the minimum among capacities of all cuts from $s$ to $t$. A cut from $s$ to $t$ is a subset of links $\tilde{\mc E} \subset \mc E$, such that $s \in \tilde{\mc E}$ and $t \notin \tilde{\mc E}$. The capacity of cut $\mc E$ is the sum of capacities of all links outgoing from $\mc E$. This equality between the solutions to the two problems is commonly referred to as the \emph{max flow min cut theorem}.

\ksmarginsoft{check the paper to what extent using $i$ for both nodes and links creates confusion}
The notion of \emph{residual} capacity will play an important role in characterization of robustness for network flow problems. Given  a flow $f \in \reals_{\geq 0}^{\mc E}$, the residual capacity of link $e \in \mc E$, node $v \in \mc V$, and of the network are, respectively, defined to be $c_e-f_e$, $\sum_{e \in \mc E_v^+} (c_e - f_e)$, and the sum of residual capacities of links outgoing from the minimum cut. The latter can be shown to be equal to the difference between the minimum cut capacity and value of $f$. 

In the presence of multiple sources and sinks, one can define a virtual super source connected via infinite capacity outgoing links to the individual sources, and similarly a virtual super sink with infinite capacity incoming links from individual sinks. One can then consider the maximum flow problem from the super source to the super sink. 


%% file: inputsJSS/notationsJSS.tex
\textbf{Input-Output Stability}: Define
$$\Ltwo^n = \theset{f : \arr_+ \rightarrow \arr_n \st \int_0^\infty f\tr(t) f(t)  \dt < \infty},$$
and
$$\LtwoE^n = \theset{f : \arr_+ \rightarrow \arr_n \st \int_0^T f\tr(t) f(t) \dt < \infty, \forAll T\in\arr_+}.$$
For $f\in \Ltwo^n$, define
$$\norm{f} = \left( \int_0^\infty f\tr(t) f(t)  \thinspace \dt \right)^{1/2}.$$
A mapping $M:\LtwoE^n \rightarrow \LtwoE^m$ is input-output stable if there exist $\alpha, \beta \ge  0$ such that
\begin{equation}
\label{eq:input-output-stable}
\norm{Mf} \le \alpha \norm{f} + \beta, \forall f\in \Ltwo^n
\end{equation}
In case $M$ is input-output stable and \textit{linear}, define
$$\norm{M} = \sup_{f\in \Ltwo^n} \frac{\norm{Mf}}{\norm{f}}$$

\textbf{Linear Dynamical Systems}: The notation
$$G \sim \sysMat{A}{B}{C}{D}$$
represents the linear time-invariant (LTI) system
\begin{align*}
\dot{x} &= Ax + Bu,\quad x(0) = 0,\\
y &= Cx + Du.
\end{align*}
with associated input-output operation
$$y(t) = D u(t) + \int_0^t Ce^{A(t-\tau)} B u(\tau) \thinspace \mathrm{d\tau}.$$
In case $A$ is a stable matrix, i.e. all eigenvalues have strictly negative real parts, then
\begin{equation}
\norm{G} = \sup_\omega \sigma_{\max} \Pmatrix{D + C(j\omega I - A)^{-1} B},
\label{eqHinf}
\end{equation}
where $\sigma_{\max}(\cdot)$ denotes the maximum singular value \cite[Chapter 4]{zhou1996robust}.

%% file: capacity-static/dc-networks-v2.tex
\section{Robustness of Finite Networks: Static Setting}
\label{sec:capacity-perturbation-static} 
Let us consider robustness of network flow to loss in link capacities. Since the quantity of interest, i.e., flow, is associated with links, for brevity in notation, we switch the indices used for nodes in previous sections to links. Accordingly, link flows, weights, and capacities will, respectively, be denoted as $f=\{f_i\}_{i \in \mc E}$, $\mc W=\{\mc W_i\}_{i \in \mc E}$ and $c=\{c_i\}_{i \in \mc E}$. 
The starting point for robustness analysis is the max flow min cut theorem for the standard static formulation, as described in Section~\ref{sec:notations}. Formally, consider the \emph{nominal} scenario where an inflow of $\lambda>0$ is routed from $s$ to $t$. $\lambda$ is less than the solution to the maximum flow problem, and hence there exists a feasible flow whose value is equal to $\lambda$. Let the link capacities be reduced by $\triangle \in [0,c]$. If $\|\triangle\|_1$ is less than the residual capacity of the unperturbed network, then there exists a new flow that is feasible for the perturbed network, i.e., it is link-wise less than $c-\triangle$ and satisfies \eqref{eq:flow-conservation}, with the same value $\lambda$. On the other hand, there exists a $\|\triangle\|_1$ infinitesimally greater than the network residual capacity, corresponding to reducing capacity on links outgoing from the minimum cut greater than their respective residual capacities, under which such a feasible flow is not possible for the perturbed network. It is of interest to extend such robustness analysis to additional constraints on flow and in presence of control.

Consider the following setup motivated by electrical networks. In order to model bi-directionality of electrical flow, while continuing to adopt the directed graph formulation of network flow from Section~\ref{sec:notations}, we do not constrain the entries of flow $f$ to be non-negative. 
In addition to \eqref{eq:flow-conservation} (Kirchhoff current law), the flow is constrained to also satisfy Ohm's law. Formally, $f \in \reals^{\mc E}$ is said to satisfy the physical constraints if there exist (voltage angles) $\phi \in \reals^{\mc V}$ such that $f=\incidence \biglambda$ and $f=\diag(\mc W) \incidence^T \phi$, where $\biglambda \in \reals^{\mc V}$ is such that $\Lambda_v$ is equal to $\lambda$ (resp., $-\lambda$) if $v$ is the source (resp., sink), and is equal to zero otherwise. This corresponds to interpreting $f$ as DC approximation to power flow, with $\mc W$ being the negative of link susceptances.  
If $\mc G$ is weakly connected, then such a unique $f$ always exists, and is given by 
\begin{equation}
\label{eq:electrical-flow}
f(\mc W,\Lambda)=\diag(\mc W) \incidence^T L^{\dagger} \Lambda
\end{equation}
 where $L^{\dagger}$ is the Moore-Penrose pseudo-inverse of the weighted Laplacian $L:=\incidence \diag(\mc W) \incidence^T$. Noting the linear dependence of $f(\mc W,\Lambda)$ on $\Lambda$, and hence $\lambda$, it is easy to see that, given $c$ and $\mc W$, the maximum flow problem in this setting can again be cast as a linear program. Let the solution be denoted as $\lambdamax(\mc W)$. 
 
 A generalization is when the link weights are flexible in a controlled manner: $\mc W \in [\Wmin, \Wmax]$, with $\Wmax \geq \Wmin \geq \zerobf$. It is then of interest to study $\max_{\mc W \in [\Wmin,\Wmax]} \lambdamax(\mc W)$, and the corresponding $\argmax$ $\mc W$. The solution to this problem can be naturally interpreted in terms of robustness to perturbation to capacity under controllable link weights, in the same spirit as the max flow min cut theorem described earlier in this section.  

\begin{figure}[htb!]
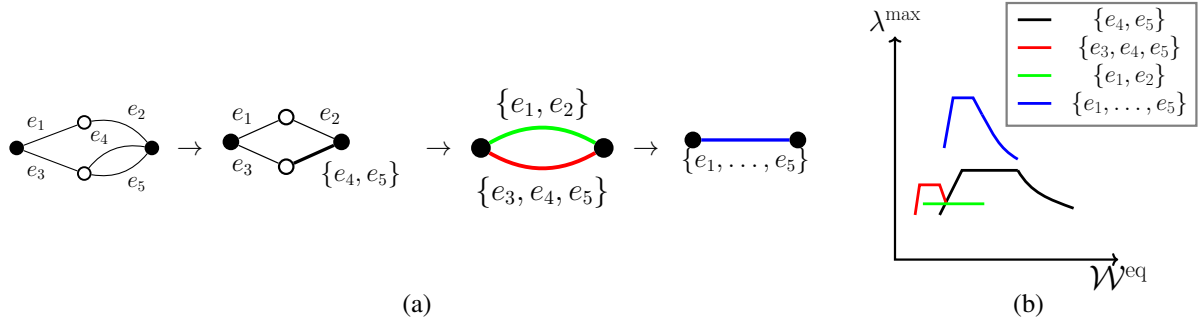

\begin{minipage}[c]{.67\textwidth}
\begin{minipage}[c]{.2\textwidth}
\begin{center}
\includestandalone[width=0.9\linewidth]{./fig/tree-reducible-net}
\end{center}
\end{minipage} $\to$
\begin{minipage}[c]{.245\textwidth}
\begin{center}
\includestandalone[width=0.9\linewidth]{./fig/tree-reduced-1}
\end{center}
\end{minipage} $\to$
\begin{minipage}[c]{.195\textwidth}
\begin{center}
\includestandalone[width=0.9\linewidth]{./fig/tree-reduced-2}
\end{center}
\end{minipage} $\to$
\begin{minipage}[c]{.185\textwidth}
\begin{center}
\vspace{0.075in}
\includestandalone[width=0.9\linewidth]{./fig/tree-reduced-3}
\end{center}
\end{minipage}
\end{minipage}
\begin{minipage}[c]{.3\textwidth}
\begin{center} \hspace{-0.3in}
\includestandalone[width=0.9\linewidth]{./fig/capacity-function-reduction-overlay}
\end{center}
\end{minipage}
\begin{minipage}[c]{.67\textwidth}
\begin{center}
(a)
\end{center}
\end{minipage}
\begin{minipage}[c]{.3\textwidth}
\begin{center}
(b)
\end{center}
\end{minipage}
\caption{(a): Steps involved in incremental network reductions of a tree reducible graph between source and sink nodes shown as solid black disks. (b): Corresponding capacity functions; for example, the capacity function for the reduced network consisting of $e_3$, $e_4$ and $e_5$ from the original network is shown in red, while the capacity function for the entire network is shown in blue.}
\label{fig:network-reduction}
\end{figure}

\begin{figure}[htb!]
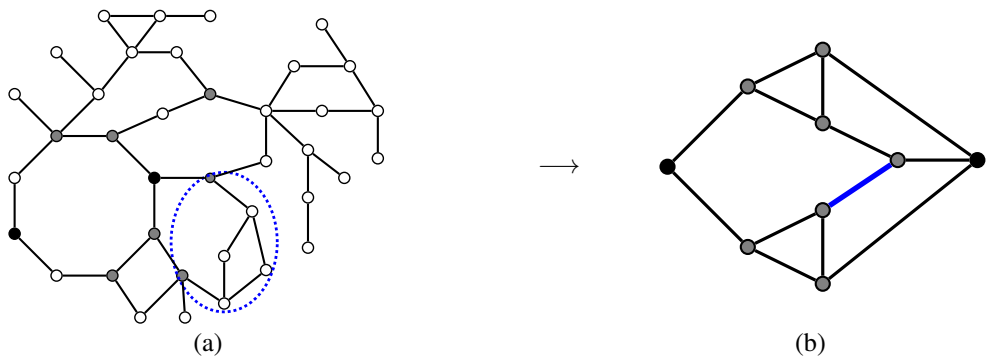

\begin{center}
\begin{minipage}[c]{.55\textwidth}
\begin{center} 
\includestandalone[width=0.55\linewidth]{./fig/ieee39-net-with-nodes-marked}
\end{center}
\end{minipage}$\longrightarrow$
\begin{minipage}[c]{.375\textwidth}
\begin{center} 
\includestandalone[width=0.7\linewidth]{./fig/ieee39-terminal-with-nodes-marked}
\end{center}
\end{minipage}

\vspace{0.05in}
\begin{minipage}[c]{.55\textwidth}
\begin{center} 
(a)
\end{center}
\end{minipage} \hspace{0.05in}
\begin{minipage}[c]{.375\textwidth}
\begin{center} 
(b)
\end{center}
\end{minipage}
\end{center}
\caption{Network reduction for the IEEE 39 benchmark network~\cite{zimmerman2011matpower}, between source and sink nodes shown as solid black disks. The original network topology is shown in (a) and the network topology after reduction is shown in (b). The original nodes which remain in the reduced network are shown as gray disks. A sample tree reducible sub-network enclosed by dashed blue ellipse in (a) is reduced to a link shown in solid blue in (b).}
\label{fig:ieee39-network-reduction}
\end{figure}

The technical challenge is due to the nonlinear dependence of $f$, and hence also of $\lambdamax$, on $\mc W$. 
\cite{Ba.Savla:TCNS16} presents an incremental network reduction approach to reduce the complexity of this problem. In the classical network flow framework (cf. Section~\ref{sec:notations}), replacing the network with a directed link from $s$ to $t$ whose capacity is the solution to the maximum flow problem is equivalent from capacity perspective. The notion of Thevenin equivalent resistance allows to replace the electrical network having link-wise weights $\mc W$ with an equivalent link with weight $\Weq$. The equivalence then maps $[\Wmin,\Wmax] \subset \reals_{\geq 0}^{\mc E}$ to, say, $[\Wmin^{\mathrm{eq}},\Wmax^{\mathrm{eq}}] \subset \reals_{\geq 0}$, and it can be shown that $\Wmin$ (resp., $\Wmax$) maps to $\Wmin^{\mathrm{eq}}$ (resp., $\Wmax^{\mathrm{eq}}$). The following rewriting:
\begin{equation}
\label{eq:max-capacity}
\max_{\mc W \in [\Wmin,\Wmax]} \lambdamax(\mc W) = \max_{\Weq \in [\Wmin^{\mathrm{eq}},\Wmax^{\mathrm{eq}}]} \lambdamax(\Weq)
\end{equation}
with $\lambdamax(\Weq):=\max\{\lambdamax(\mc W): \, \mc W \in [\Wmin, \Wmax] \text{ s.t. equivalent weight of $\mc W$ is } \Weq\}$ suggests $\lambdamax(\Weq)$ as the capacity associated with $\Weq$. This notion of capacity lends itself to the following iterative solution of \eqref{eq:max-capacity} if $\mc G$ is \emph{tree reducible}, i.e., if $\mc G$ can be reduced to a single link by a sequence of series and parallel subnetwork reductions, as illustrated in Figure~\ref{fig:network-reduction}. \cite{Ba.Savla:TCNS16} shows that a particular quasi-concave property of capacity functions $\lambdamax(.)$ remains invariant across each of these reduction steps, see Figure~\ref{fig:network-reduction}(b).  Moreover, the interval over which the capacity function for a network achieves its maximum can be analytically related to the corresponding intervals of its sub-networks, thereby allowing an analytical solution to \eqref{eq:max-capacity}. In other words, this procedure provides an analytical solution to the non-convex problem in \eqref{eq:max-capacity} for tree reducible networks. For other $\mc G$, one can perform this reduction for each of tree-reducible sub-networks of $\mc G$, and thereby reducing computational complexity; see Figure~\ref{fig:ieee39-network-reduction} for an illustration.

\paragraph{Future Research Directions} 
Extensions to other physical constraints, e.g., in AC power flow, natural gas and water networks, along with relevant control actions, are natural directions to pursue. It is also of interest to pursue alternate techniques to tackle the resulting non-convexity in capacity and robustness analysis. Several interesting properties of the optimization problem in \eqref{eq:max-capacity}, such as sufficient conditions for equivalence of local and global optimal solutions,  as provided in \cite[Chapter 3]{Ba:PhDThesis}, suggest possible directions.

%% file: fig/tree-reducible-net.tex
\begin{tikzpicture}[node distance=0.4cm and 1.5cm]
 \node[supply node] (1) {};
 \node[main node] (2) [above right =of 1] {};
 \node[main node] (3) [below right =of 1] {};
 \node[demand node] (4) [below right =of 2] {};

 \path[main edge, edge label]
   (1) edge node{\LARGE $e_1$} (2)
   (1) edge node[below left]{\LARGE $e_3$} (3)
   (2) edge[bend left] node{\LARGE $e_2$} (4)
   (3) edge[bend left] node {\LARGE $e_4$} (4)
   (3) edge[bend right] node[below right] {\LARGE $e_5$} (4);

\end{tikzpicture}

%% file: fig/tree-reduced-1.tex
\begin{tikzpicture}[node distance=0.3cm and 1cm]
 \node[supply node] (1) {};
 \node[main node] (2) [above right =of 1] {};
 \node[main node] (3) [below right =of 1] {};
 \node[demand node] (4) [below right =of 2] {};

 \path[main edge, edge label]
      (1) edge node{\LARGE $e_1$} (2)
     (2) edge node{\LARGE $e_2$} (4)
    (1) edge node[below left]{\LARGE $e_3$} (3)
   (3) edge[line width=0.75mm] node[below right] {\LARGE $\{e_4,e_5\}$} (4);
\end{tikzpicture}

%% file: fig/tree-reduced-2.tex
\begin{tikzpicture}[node distance=2cm]
 \node[supply node] (1) {};
 \node[demand node] (4) [right =of 1] {};

 \path[main edge, edge label]
   (1) edge[bend left, green, line width = 0.75mm] node{\color{black}\LARGE $\{e_1,e_2\}$} (4)
   (1) edge[bend right, red, line width = 0.75mm] node[below]{\color{black}\LARGE $\{e_3,e_4,e_5\}$} (4);

\end{tikzpicture}

%% file: fig/tree-reduced-3.tex
\begin{tikzpicture}[node distance=2cm]
 \node[supply node] (1) {};
 \node[demand node] (4) [right =of 1] {};

 \path[main edge, edge label]
   (1) edge[blue, line width=0.75mm] node[below]{\color{black}\LARGE $\{e_1,\ldots, e_5\}$} (4);

\end{tikzpicture}

%% file: fig/capacity-function-reduction-overlay.tex
\begin{tikzpicture}
\draw[<->, thick, line width =1mm] (0,10) node [above,scale=1.5,black] {\Huge $\lambdamax$}-- (0,0) node [below left] {} -- (10,0) node [below,scale=1.75,black] {\Huge $\Weq$}; 

\draw[line width=1.25mm,black] (2, 2) to (3, 4) to (5.5, 4); 
\draw[line width=1.25mm,black] (5.5, 4) [out = -60, in = 160] to (8,2.3);


\draw[line width=1.25mm,red] (0.9, 2) to (1.1, 3.35) -- (2,3.35) -- (2.3,2.5);
\draw[line width=1.25mm,green] (1.25, 2.5) to (4, 2.5);


\draw[line width=1.25mm,blue] (2.2, 5) to (2.6, 7.25) -- (3.5,7.25) -- (3.9,6.5);
\draw[line width=1.25mm,blue] (3.9,6.5) [out = -60, in = 150] to (5.5,4.5);


\draw[line width=1.25mm,blue] (5.5,7) -- (7,7);
\draw (10.5,7) node [scale=1.25,black] {\Huge $\{e_1, \ldots, e_5\}$};
\draw[line width=1.25mm,green] (5.5,8.25) -- (7,8.25);
\draw (10.5,8.25) node [scale=1.25,black] {\Huge $\{e_1, e_2\}$};
\draw[line width=1.25mm,red] (5.5,9.5) -- (7,9.5);
\draw (10.5,9.5) node [scale=1.25,black] {\Huge $\{e_3, e_4, e_5\}$};
\draw[line width=1.25mm,black] (5.5,10.75) -- (7,10.75);
\draw (10.5,10.75) node [scale=1.25,black] {\Huge $\{e_4, e_5\}$};

\draw [draw=gray,line width=1mm] (5,6) rectangle (13.5,11.5);

\end{tikzpicture}

%% file: fig/ieee39-net-with-nodes-marked.tex
\begin{tikzpicture}                   
 \node[supply node] (39) at (0, -1) {$\text{  }$};   \node[main node] (1) at (0, 1) {$\text{  }$};
 \node[main node, fill=gray] (2) at (1.5, 2.5) {$\text{  }$};  \node[main node] (9) at (1.5, -2.5) {$\text{  }$};
  \node[main node, fill=gray] (3) at (3.5, 2.5) {$\text{  }$};   \node[main node, fill=gray] (8) at (3.5, -2.5) {$\text{  }$}; 
  \node[demand node] (4) at (5, 1) {$\text{  }$};   \node[main node, fill=gray] (5) at (5, -1) {$\text{  }$}; 
  \node[main node, fill=gray] (6) at (6, -2.5) {$\text{  }$};     \node[main node] (7) at (4.5, -4) {$\text{  }$};    \node[main node] (31) at (6.1, -4) {$\text{  }$}; 
  \node[main node, fill=gray] (14) at (7, 1) {};      \node[main node] (13) at (8.5, -0.2) {$\text{  }$}; 
   \node[main node] (11) at (7.5, -3.5) {$\text{  }$};
 \node[main node] (12) at (7.5, -1.8) {$\text{  }$};  \node[main node] (10) at (9, -2.3) {$\text{  }$}; 
 \node[main node] (18) at (5.3, 3.3) {$\text{  }$};  \node[main node, fill=gray] (17) at (7, 4) {$\text{  }$}; 
   \node[main node] (15) at (9, 1.6) {$\text{  }$};     \node[main node] (16) at (9, 3.4) {$\text{  }$};  
   
    \node[main node] (25) at (3, 4) {$\text{  }$};  
    \node[main node] (26) at (4.2, 5.5) {$\text{  }$};   \node[main node] (27) at (5.8, 5.5) {$\text{  }$};    
         \node[main node] (30) at (0, 4) {$\text{  }$};      \node[main node] (37) at (1.5, 5.5) {$\text{  }$}; 
     \node[main node] (28) at (3.2, 6.8) {$\text{  }$};      \node[main node] (29) at (5.2, 6.8) {$\text{  }$};  
     \node[main node] (38) at (7, 6.8) {$\text{  }$}; 
     
     \node[main node] (24) at (10, 5) {$\text{  }$};       \node[main node] (23) at (12, 5) {$\text{  }$};    \node[main node] (36) at (11, 6.5) {$\text{  }$};
     \node[main node] (21) at (11, 3.4) {$\text{  }$};   \node[main node] (22) at (13, 3.4) {$\text{  }$};    \node[main node] (35) at (13, 1.7) {$\text{  }$};
     \node[main node] (19) at (10.5, 2) {$\text{  }$};        \node[main node] (33) at (11.8, 1) {$\text{  }$};
      \node[main node] (20) at (10.5, 0.3) {$\text{  }$};     \node[main node] (34) at (10.5, -1.5) {$\text{  }$};

\draw[dashed, thick, blue,line width = 1mm] (7.5,-1.3) ellipse (1.9 and 2.5) ;

\path[cascading edge]
	(1) edge (2)        
	(1) edge  (39)     
	(2) edge (3)	     
	(2) edge (30)	     
	(2) edge (25)	     
	(3) edge (4)	     
	(3) edge (18)	     
	(4) edge (5)  	
	(4) edge (14) 	 
	(5) edge (6)  	 
	(5) edge (8)  	
	(6) edge (7)   	
	(6) edge (11)  	 
	(6) edge (31)   	
	(7) edge (8)   	
	(8) edge  (9)      
	(9) edge (39)      
	(10) edge (11)   
	(10) edge (13)   
	(11) edge (12)   
	(12) edge (13)   
	(13) edge (14)   
	(14) edge (15)   
	(15) edge  (16) 		
	(16) edge  (19) 		
	(16) edge  (21) 		
	(16) edge  (24) 		
	(17) edge  (16) 		
	(17) edge  (18) 		
	(17) edge (27)	     
	(19) edge  (20) 		
	(19) edge  (33) 		
	(20) edge  (34) 		
	(21) edge  (22) 		
	(22) edge  (23) 		
	(22) edge  (35) 		
	(23) edge  (36) 		
	(23) edge  (24) 		
	(25) edge (26)	     
	(25) edge (37)	     
	(26) edge (27)	     
	(26) edge (28)	     
	(26) edge (29)	     
	(28) edge (29)	     
	(29) edge (38)	     
	;     	
\end{tikzpicture} 

%% file: fig/ieee39-terminal-with-nodes-marked.tex
\begin{tikzpicture}[node distance=0.6cm and 1.5cm]
 \node[supply node] (39) {};
 \node[main node, fill=gray] (2) [above right =2.3cm of 39] {};
 \node[main node, fill=gray] (3) [above right =of 2] {};
 \node[main node, fill=gray] (17) [below right =of 2] {};

 \node[main node, fill=gray] (8) [below right =2.3cm of 39] {};
 \node[main node, fill=gray] (5) [below right =of 8] {};
 \node[main node, fill=gray] (6) [above right = of 8] {};
 
 \node[main node, fill=gray] (14) [below right = of 17] {};
 \node[demand node] (4) [right = of 14] {};

 \path[cascading edge, edge label]
   (39) edge  (2) 
   (39) edge  (8)
   (2) edge  (3) 
   (2) edge  (17)
   (3) edge  (17) 
     
    (8) edge (5)   
    (8) edge  (6)   
    (5) edge (6)   

    (17) edge  (14)   
    (6) edge[blue, line width=1.25mm] (14) 
    
    (3) edge  (4) 
    (14) edge (4) 
    (5) edge (4) ;
      
\end{tikzpicture}

%% file: non-asymptotic-dynamic/non-asymptotic-dynamic-main-v2.tex
\section{Robustness of Finite Networks: Linear Dynamical Setting}
\label{sec:non-asymptotic-dynamical}

We provide background on robustness analysis tools in Sections~\ref{sec:M-Delta} and \ref{sec:stochastic-stability-new}, and discuss application to network dynamics in Section~\ref{sec:applications} through an illustrative example. 

\input{./inputsJSS/mDeltaJSS}

\subsection{Memoryless Stochastic Structured Uncertainties}
\label{sec:stochastic-stability-new}
Consider \eqref{eqSmallGain} with $d$ and $\omega$ as white second order processes, and  
\begin{equation}
\label{eq:Delta-structured-uncertainty}
\bigtriangle(t)  = \begin{bmatrix} \triangle_1(t) & & \\ & \triangle_2(t) & \\ & \ldots & \\ & & 
\triangle_m(t) \end{bmatrix}
\end{equation}
is a diagonal matrix of time-varying gains $\{\triangle_i\}_{i \in [m]}$, modeled as zero mean random processes that are temporally independent, but possibly mutually correlated ($m$ is not necessarily the dimension of state space realization of $M$ in \eqref{eqSmallGain}). 
The entries of $d$ can also possibly be mutually correlated, and so can the entries of $\omega$. For simplicity, here we let all the quantities be real valued and let the covariance matrices of $d$ and $\omega$ be time-invariant; please refer to \cite{Bamieh.Filo:18} for extensions.

In the above setting, the feedback system in \eqref{eqSmallGain} (cf. Figure~\ref{figSmallGain}) is called mean-square (MS) stable if signals $y$ and $r$ have uniformly bounded variance sequences, i.e., if there exists a constant $c$ such that
$$
\max \left\{ \|\E[y(t)y^T(t)]\|_{\infty}, \|\E[r(t)r^T(t)]\|_{\infty} \right\} \leq c 
$$


The next result~\cite[Theorem 3.2 and Section VIC]{Bamieh.Filo:18} gives a tight condition on MS stability. In preparation for the result, let $\Sigma_{\bigtriangle}$ and $\Sigma_{\omega}$ be, respectively, the covariance matrices of $[\triangle_1(t) \, \ldots \triangle_m(t)]^T$ and $\omega$, and let $\{M_t\}$ be the matrix-valued impulse response sequence of $M$. 

\begin{theorem}
\label{thm:stable-main}
Let $M$ be a stable (finite $\mc H^2$ norm), strictly-causal, LTI system.  \eqref{eqSmallGain} and \eqref{eq:Delta-structured-uncertainty} is MS stable if and only if $\rho(\Hop)<1$, where
the matrix-valued linear operator $\Hop$, also known as the \emph{loop gain operator}, is defined as:
\begin{equation*}
\label{eq:H-def}
\begin{split}
\Hop(X) :=\Sigma_{\bigtriangle} \circ \left(\sum_{t=0}^{\infty} M_t X M_t^T \right)
\end{split}
\end{equation*}
If $\rho(\Hop)>1$ and $\Sigma_{\omega}$ is equal to the \emph{Perron eigen-matrix} of $\Hop$, then the covariance $\E[u_t u_t^T]$ grows unbounded geometrically. 
%
\end{theorem}

\cite[Section IVA]{Bamieh.Filo:18} establishes that $\rho(\|M\|_2^2)<1$ is also a necessary and sufficient condition for MS stability of \eqref{eqSmallGain} and \eqref{eq:Delta-structured-uncertainty} when $\Sigma_{\triangle}=I$ (uncorrelated uncertainties), where $\|M\|_2^2$ is the matrix of squared $\mc H^2$ norms of subsystems of $M$. However, for correlated uncertainties, i.e., $\Sigma_{\triangle} \neq I$, \cite[Section IVC]{Bamieh.Filo:18} argues that such a condition involving only $\mc H^2$ norms of subsystems is not enough, and that one needs in addition other system metrics, like the inner product between different subsystems' impulse responses. 

The tight stochastic stability result in Theorem~\ref{thm:stable-main} is to be contrasted with techniques that approximate stochastic linear dynamics, in state-space form, using polynomial chaos expansions, and then project into a higher dimensional \emph{deterministic} linear dynamics. Standard Lyapunov argument is then used to analyze stability of the approximate deterministic system, e.g., see \cite{Fisher.Bhattacharya:09,Kim.Shen.ea:13}. On the other hand, these approximate techniques are applicable to more general parametric uncertainties.

\cite[Section VI]{Bamieh.Filo:18} suggests that it might be convenient to compute $\rho(\Hop)$ using state space representation. If 
\begin{equation}
\label{eq:M-state-space}
M \sim
\left(
\begin{array}{c|c}
\Azero & B \\
\hline
C & 0
\end{array}
\right)
\end{equation}
then $\rho(\Hop)$ is equal to the largest real number $\lambda$ such that the following LMI has a feasible solution $X \geq 0$:
\begin{equation*}
\label{eq:LMI}
\lambda \left(X - \Azero X {\Azero}^T \right) - B \left(\Sigma_{\bigtriangle} \circ (C X C^T)\right) B^T = 0 
\end{equation*}

\subsection{Application to Network Dynamics with Unreliable Links} 
\label{sec:applications}
The $M-\triangle$ framework in Section~\ref{sec:M-Delta} and the associated small gain techniques can be applied in the network context to handle specifics of where model perturbations occur in a feedback interconnection. An illuminating setup is in state-space form such as:

\begin{equation}
\label{eq:dynamics-mult-uncertainty}
x(t+1) = \left( \Azero + \sum_{j \in [m]} \triangle_j(t) A^{(j)} \right) x(t) + \omega(t)
\end{equation}
where $A^{(0)}$ is interpreted as the nominal system description and the parametric uncertainty in the nominal system is reflected by the $m$ scalar quantities $\{\triangle_j\}_{j \in [m]}$. The structural knowledge about the uncertainty is contained in $\{A^{(j)}\}_{j \in [m]}$. The form of \eqref{eq:dynamics-mult-uncertainty} can be converted into \eqref{eqSmallGain} with the structure of $\bigtriangle$ as in \eqref{eq:Delta-structured-uncertainty} \cite{zhou1996robust}.

\input{./non-asymptotic-dynamic/examples-v2}



%
%


%% file: inputsJSS/mDeltaJSS.tex
\subsection{Arbitrary Structured Uncertainties}
\label{sec:M-Delta}
\begin{figure}
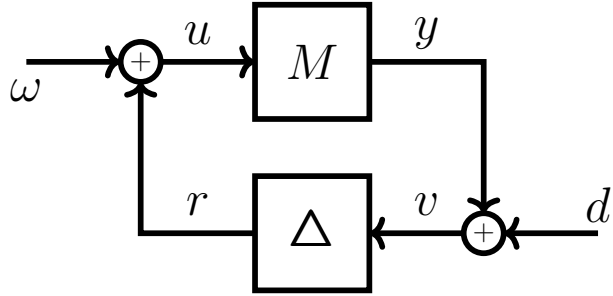

\begin{center}
\includestandalone[width=0.5\textwidth]{./fig/M-Delta-v2}
\caption{Robustness analysis configuration.}
\label{figSmallGain}
\end{center}
\end{figure}

The question of robust stability concerns whether a dynamical system maintains stability in the presence of a specified \textit{family} of perturbations. 

A well studied setting is illustrated in Figure~\ref{figSmallGain}, which represents the feedback equations
\begin{subequations}\label{eqSmallGain}
\begin{align}
u &= \omega + \Delta v \, \kscommentjss{ =: \omega + r}\\
v &= d + Mu \, \kscommentjss{ =: d + y}
\end{align}
\end{subequations}
We assume that these equations are well-posed in that for all $w, d\in \LtwoE^n$, there exist unique $u,v\in\LtwoE^n$ satisfying (\ref{eqSmallGain})\footnote{A further technical assumption is that the implied mapping $(w,d)\mapsto \kscommentjss{(u,v)}$ is \textit{causal} \cite{willems1970analysis}.}. 
Let $T[M,\Delta]$ denote the implied closed-loop input-output mapping, i.e.,
$$\Pmatrix{u\\ v} = T[M,\Delta] \Pmatrix{w\\ d}.$$

The setup in Figure~\ref{figSmallGain} is as follows:
\begin{itemize}
\item $M:\LtwoE^n \rightarrow \LtwoE^n$ is an input-output stable mapping satisfying
$$\norm{Mf} \le \alpha_M \norm{f} + \beta_M,\forall f\in \LtwoE^n.$$
\item $\Delta \in \mathbf{\Delta}$, where $\mathbf{\Delta}$ represents a \textit{family} of input-output stable mappings satisfying
$$\mathbf{\Delta} = \theset{ \Delta :  \LtwoE^n \rightarrow \LtwoE^n \st \norm{\Delta f} \le \alpha_\delta \kscommentjss{\norm{f}} + \beta_\delta.}$$
\end{itemize}

For a specific $\Delta\in\mathbf{\Delta}$, the feedback system (\ref{eqSmallGain}) is \textit{closed-loop input-output stable} if $T[M,\Delta]$ is input-output stable, i.e., there exist $\alpha, \beta \ge 0$ such that
$$\norm{\Pmatrix{u\\ v}} = \norm{T[M,\Delta]\Pmatrix{w\\ d}}\le \alpha \norm{\Pmatrix{w\\ d}} + \beta.$$
The feedback system (\ref{eqSmallGain}) is \textit{robustly stable} with respect to $\mathbf{\Delta}$ if $T[M,\Delta]$ is input-output stable for \textit{all} $\Delta\in\mathbf{\Delta}$.

Our starting point is the classical small-gain theorem \cite{desoer1975feedback}. See  \cite{jiang2018small} for a recent survey.

\begin{theorem}
The closed-loop system (\ref{eqSmallGain}) is robustly stable with respect to $\mathbf{\Delta}$ if
$$\alpha_M \alpha_\delta < 1.$$
\end{theorem}

Note that the small gain condition as stated is only a sufficient condition for robust stability. Also of interest is when the small gain condition is necessary. 

For this discussion, we will restrict our attention to \textit{linear system models} for the remainder of this section\footnote{A nonlinear setting where a small gain condition is necessary is where $M$ has \textit{fading memory}  \cite{shamma1991necessity,freeman2001necessity}.}. Define 
$$\mathbf{\Delta}_\mathrm{LTI}  = \theset{\Delta :  \LtwoE^n \rightarrow \LtwoE^n \st \Delta \text{ is linear time-invariant} \And \norm{\Delta} \le 1}.$$

\begin{theorem}[Theorem 9.1 \cite{zhou1996robust}]\label{thmLTIff} Let $M$ be a linear time-invariant system. The closed-loop system (\ref{eqSmallGain}) is robustly stable with respect to $\mathbf{\Delta}_\mathrm{LTI}$ if and only if
$$\norm{M} < 1.$$
\end{theorem}
The meaning of the small gain condition being necessary is that if the small gain condition is violated, i.e., $\norm{M} \ge 1$,  then there exists an admissible $\Delta\in\mathbf{\Delta}_\mathrm{LTI}$ resulting in a closed loop $T[M,\Delta]$ that is not stable. An explicit construction is provided in 
\cite[Theorem 9.1]{zhou1996robust}. The general idea is as follows. Suppose that $M \sim \sysMat{A}{B}{C}{D}$ and
$$\norm{M} \ge 1.$$ 
Then, recalling (\ref{eqHinf}), there exists an $\omega^*$ such that
\begin{equation}\label{eqBigSig}
\sigma_{\max}\Pmatrix{D + C(j\omega^* I - A)^{\kscommentjss{-1}}B} \ge 1.
\end{equation}
Define
$$\hat{M}(j\omega^*) = D + C(j\omega^* I - A)^{\kscommentjss{-1}}B.$$
A consequence of (\ref{eqBigSig}) is that
there exists a $n\times n$ complex matrix $Q$ with $\sigma_{\max}(Q) \le 1$ such that
\begin{equation}\label{eqBadQ}
\mathrm{det}(I - Q\hat{M}(j\omega^*)) = 0.
\end{equation}
Finally, one can construct an admissible $\Delta^*\in \mathbf{\Delta}$ with representation $\sysMat{A_\delta}{B_\delta}{C_\delta}{D_\delta}$ such that
$$D_\delta + C_\delta (j\omega^* I - A_\delta)^{-1} B_\delta = Q.$$
The implication is that the closed-loop system will be unstable, in particular with a pole at $j\omega^*$, and $\Delta^*$ is a destabilizing perturbation.

\begin{figure}
\begin{center}
\includegraphics[width=0.5\textwidth]{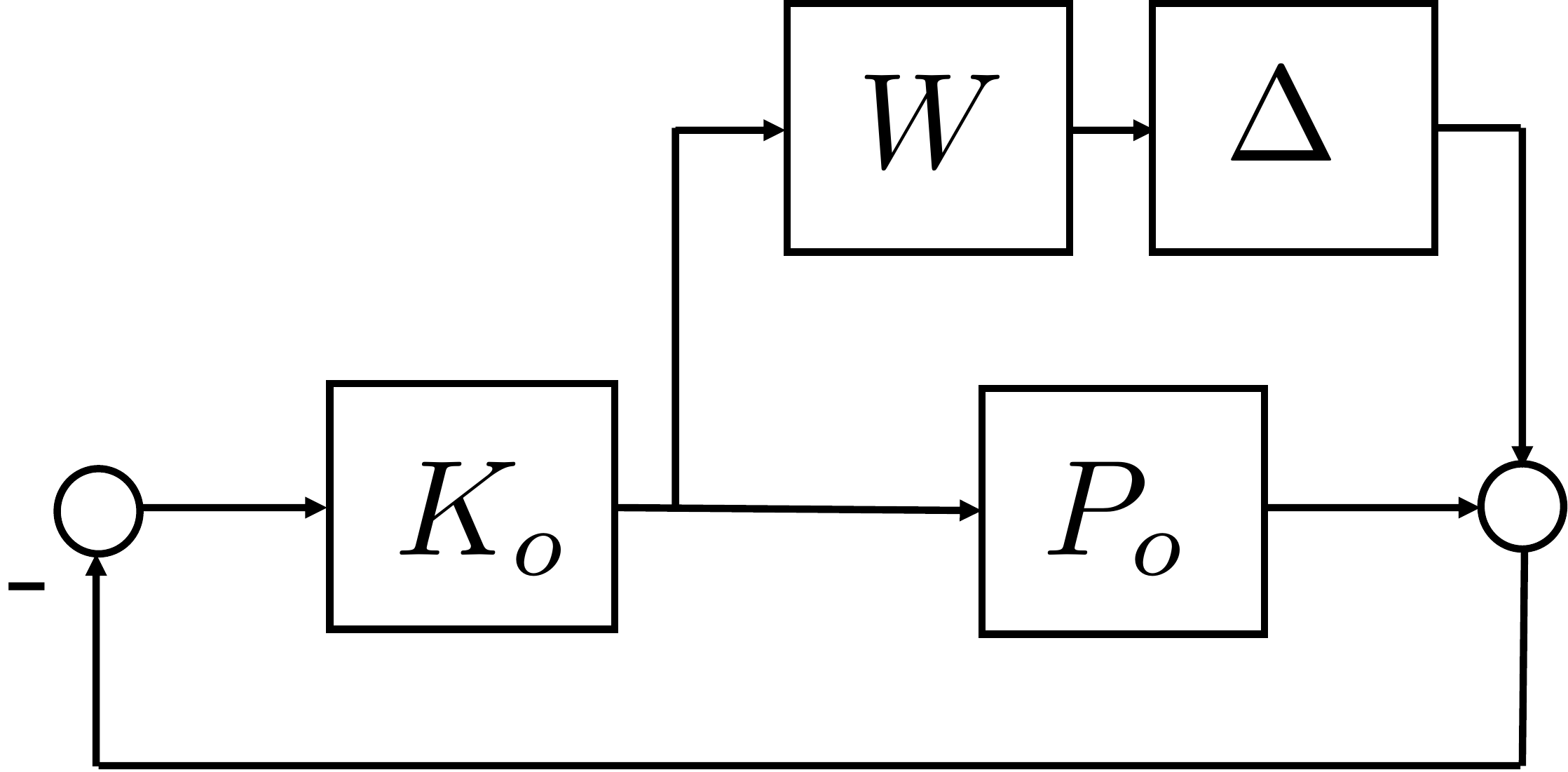}
\caption{Additive error acting on nominal plant model, $P_o$.}
\label{figAdditive}
\end{center}
\end{figure}

The feedback configuration in Figure~\ref{figSmallGain} is constructed by isolating the effects of modeling errors. An illustrative scenario is in Figure~\ref{figAdditive}. In this setup, $P_o$ is a nominal plant model to be controlled by $K_o$. However, the family of possible plant models is 
$P_o + \Delta W$, where the system $W$ acts a dynamic weighting on the effects modeling error $\Delta\in \mathbf{\Delta}_\mathrm{LTI}$. Transforming Figure~\ref{figAdditive} to Figure~\ref{figSmallGain} results in
$$M = -WK_o(I+P_oK_o)^{-1}.$$
Accordingly, a necessary and sufficient condition for robust stability is that
$$\norm{WK_o(I+P_oK_o)^{-1}} < 1.$$

\begin{figure}
\begin{center}
\includegraphics[width=0.7\textwidth]{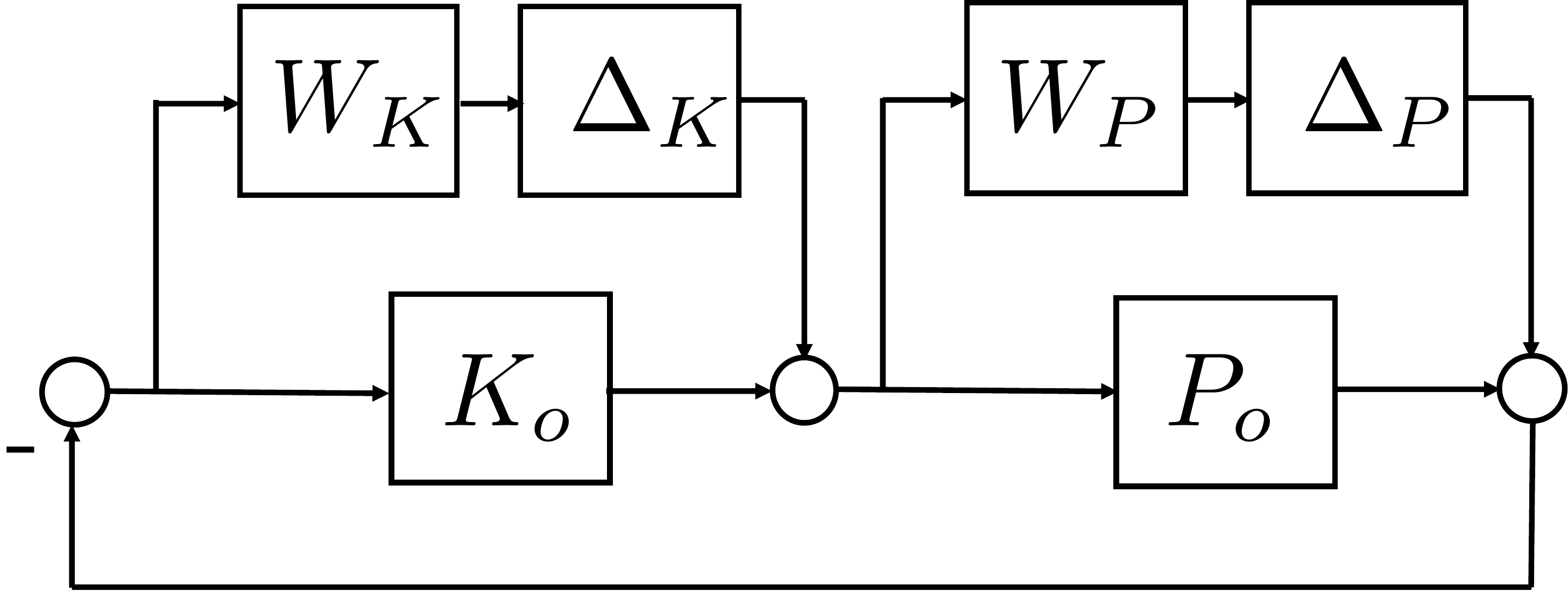}
\caption{Additive error acting on both a nominal plant model, $P_o$, and nominal controller, $K_o$.}
\label{figAdditiveX2}
\end{center}
\end{figure}

Now consider the scenario illustrated in Figure~\ref{figAdditiveX2}.  For convenience, let us assume that all mappings are single-input/single output.
As before, $P_o$ is a nominal plant model with weighted additive error. However, there is now weighted additive error impacting the nominal controller, $K_o$, as well. Transforming Figure~\ref{figAdditiveX2} to the small gain configuration of Figure~\ref{figSmallGain} results in
\begin{equation}\label{eqAdditiveX2}
M = \Pmatrix{-W_P K_o (I+P_oK_o)^{-1}& W_P (I+K_oP_o)^{-1}\\ -W_K(I+P_oK_o)^{-1}& -W_KP_o(I+K_oP_o)^{-1}}
\end{equation}
and
\begin{equation}\label{eqAdditiveX2deltas}
\Delta = \Pmatrix{\Delta_P & 0\\ 0 & \Delta_K}.
\end{equation}
We see that in the case where there is a graphical structure on the location of model perturbations, the resulting transformation to the form of Figure~\ref{figSmallGain} results in an associated restricted structure in $\Delta$.

To analyze the consequences,  define $\mathbf{\Delta}_\mathrm{LTI}^\mathrm{diag} \subset \mathbf{\Delta}_\mathrm{LTI}$ to be the subset of \textit{diagonal} stable LTI systems with norm less than one.
While $\norm{M} < 1$ remains a sufficient condition for robust stability with respect to $\mathbf{\Delta}_\mathrm{LTI}^\mathrm{diag}$, it is no longer the case that this condition is necessary. As before, let us assume that $\norm{M} \ge 1$ and that condition (\ref{eqBigSig}) holds. As before, it is still possible to construct a complex matrix $Q$ with $\sigma_{\max}(Q) \le 1$ such that (\ref{eqBadQ}) holds. However, this $Q$ need not be \textit{diagonal}, and so it may not be possible to construct an admissible destabilizing $\Delta\in \mathbf{\Delta}_\mathrm{LTI}^\mathrm{diag}$ as before. In terms of controller synthesis, seeking $\norm{M} < 1$ is unnecessarily restrictive.

The following special case of rank-one matrices illustrates the main idea. Define $\mathbf{Q}^\mathrm{full}$ to be the set of
complex $n \times n$ matrices, and let $\mathbf{Q}^\mathrm{diag} \subset \mathbf{Q}^\mathrm{full}$ denote the subset of diagonal matrices. 

\begin{proposition} Let $X = ab^\mathrm{H}$ be a rank-one matrix, defined by column vectors $a,b \in \mathbf{C}_n$.
\begin{enumerate}
\item $\mathrm{det}(I - Q X) \not= 0$ for all $Q\in \mathbf{Q}^\mathrm{full}$ with $\sigma_\mathrm{max}(Q) \le 1$ if and only if $\sigma_{\max}(X) = \norm{a}\norm{b} < 1$.
\item $\mathrm{det}(I - QX) \not= 0$ for all $Q\in \mathbf{Q}^\mathrm{diag}$ with $\sigma_\mathrm{max}(Q) \le 1$ if and only if $\sum_{i=1}^n \magn{a_i}\magn{b_i} < 1$.
\end{enumerate}
\end{proposition}

Since in general
$$\sum_{i=1}^n \magn{a_i}\magn{b_i} \le \norm{a}\norm{b},$$
the restriction to diagonal $Q$ matrices implies that larger $X$ that fail Condition 1 can still satisfy Condition 2. Indeed, depending on the specific structure of $a$ and $b$, the difference  can be significant.

These considerations led to the introduction of the structured singular value \cite{doyle1982analysis,safonov1982stability}. See \cite{packard1993complex} and references therein for an extensive discussion. The definition is as follows. Let $\mathbf{Q}$ denote a class of complex $n\times n$ matrices, where $\mathbf{Q}^\mathrm{full}$ and $\mathbf{Q}^\mathrm{diag}$ are specific examples. Now define\footnote{$\sigma_\mathrm{ssv}(X;\mathbf{Q}) = 0$ whenever $\mathrm{det}(I-QX) = 0$ is not possible
for any $Q\in \mathbf{Q}$.}
$$\sigma_\mathrm{ssv}(X;\mathbf{Q}) = \frac{1}{
\inf \theset{\sigma_{\max}(Q) \st Q\in \mathbf{Q} \And \mathrm{det}(I-QX) = 0}}.$$
Note that $\sigma_\mathrm{ssv}(\cdot ; \mathbf{Q})$ must be defined with respect to a family of matrices. In the case of $\mathbf{Q}^\mathrm{full}$, then for any matrix, $X$, the structured singular value defaults to the standard singular value, i.e., 
$$\sigma_\mathrm{ssv}(X; \mathbf{Q}^\mathrm{full}) = \sigma_{\max}(X).$$

By construction, the structured singular value can be used to derive necessary and sufficient conditions for robust stability in the case of structured perturbations. A lingering issue is its efficient computation. It is possible to compute an upper bound that is sometimes exact. The following result is representative.

\begin{theorem}[\cite{doyle1982analysis}] Let $X$ be an $n\times n$ complex matrix. Then
$$\sigma_\mathrm{ssv}(X; \mathbf{Q}^\mathrm{diag}) \le \inf_{D > 0,\ \mathrm{diagonal}} \sigma_{\max}( DXD^{-1}).$$
Furthermore, in the case of $n\le 3$, equality holds.
\end{theorem}

There is extensive literature addressing variations, extensions, and computational analysis of the main results of this section, including the case of slowly varying perturbations \cite{poolla1995robust}, time-varying perturbations \cite{megretski1993necessary,shamma1994robust}, and alternative system norms \cite{khammash1991performance}. Furthermore, the discussion herein presented a small-gain approach, which is a special case of dissipativity analysis \cite{willems2007dissipative}. See \cite{arcak2016networks} for an extensive discussion of dissipativity for networked systems.

\begin{figure}
\begin{center}
\includegraphics[width=0.5\textwidth]{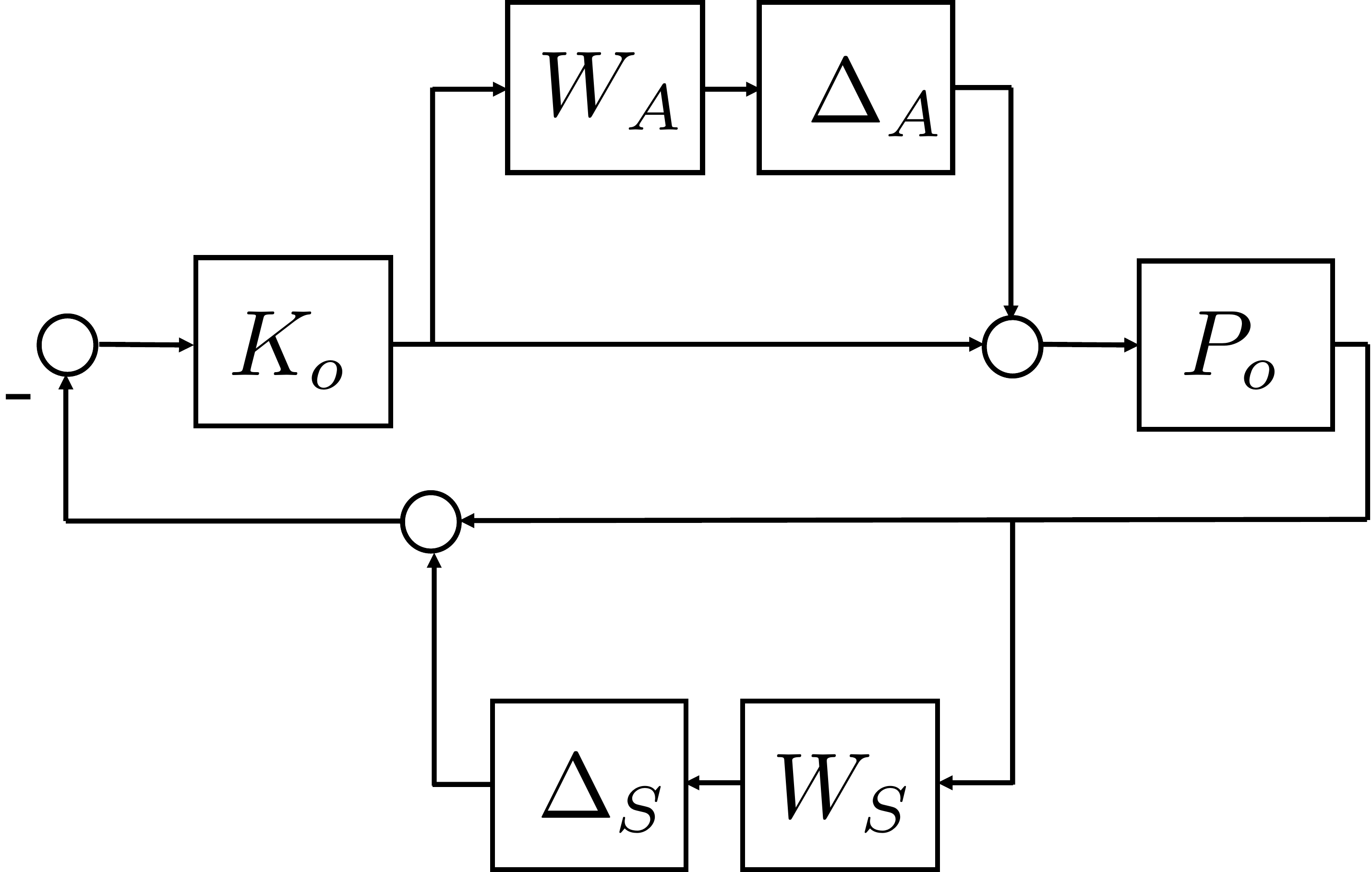}
\caption{Modeling errors in the actuation and sensing channels.}
\label{figMultiplicativeX2}
\end{center}
\end{figure}

The main emphasis here is the implications of robustness analysis when there is an underlying graphical structure. To reinforce this point, consider now the scenario illustrated in Figure~\ref{figMultiplicativeX2}. This diagram models feedback control in the presence of modeling errors in the actuation ('$A$') forward loop and sensing ('$S$') feedback loop, both of which can be viewed as a consequence of controlling a system over a network with dynamic channels. Transforming Figure~\ref{figMultiplicativeX2} to the small gain configuration of Figure~\ref{figSmallGain} results in
\begin{equation}\label{eqMultiplicativeX2}
M = \Pmatrix{-W_AK_oP_o(I+K_oP_o)^{-1}& -W_AK_o(I+P_oK_o)^{-1}\\
W_S P_o(I+K_oP_o)^{-1}& -W_SP_oK_o(I+P_oK_o)^{-1}}
\end{equation}
and
\begin{equation}\label{eqMultiplicativeX2deltas}
\Delta = \Pmatrix{\Delta_A & 0\\ 0&\Delta_S}.
\end{equation}
In comparing the two scenarios of Figure~\ref{figAdditiveX2} and Figure~\ref{figMultiplicativeX2} when transformed to the standard configuration of Figure~\ref{figSmallGain}, we see that the analysis differs in comparing the resulting $M$ in (\ref{eqAdditiveX2}) versus in (\ref{eqMultiplicativeX2}). The resulting $\Delta$ family in (\ref{eqAdditiveX2deltas}) versus (\ref{eqMultiplicativeX2deltas}) are effectively the same, despite the different origins. Accordingly, the robustness conclusions depend on the underlying graphical structure of where uncertainty enters into the overall system.

In closing this section, we also mention relatively later developments in robustness analysis which use the \emph{integral quadratic constraints} (IQC) framework~\cite{Megretski.Rantzer:97}. Enabled by advancements in related computational algorithms, e.g., see \cite{Nesterov.Nemirovskii:94,Boyd.ElGhaoui.ea:94}, these techniques allow to efficiently extend robustness analysis to several canonical nonlinearities and time-varying uncertainties in the $\bigtriangle$ block.

%% file: fig/M-Delta-v2.tex
\begin{tikzpicture}
\draw[very thick,line width=1mm] (-1,-1) rectangle (1,1);
\draw[very thick] (0,0) node {\Huge $M$};

\draw[very thick,line width=1mm] (-1,-4) rectangle (1,-2);
\draw[very thick] (0,-3) node {\Huge $\bigtriangle$};

\draw[very thick,->,line width=1mm] (1,0) -- (3,0) -- (3,-3+0.35); 
\draw[very thick,->,line width=1mm] (3-0.35,-3) -- (1,-3);

\draw[very thick,line width=1mm] (-3,0) circle(0.35);
\draw[very thick,line width=1mm] (3,-3) circle(0.35);

\draw[very thick,->,line width=1mm] (-3+0.35,0) -- (-1,0);
\draw[very thick,->,line width=1mm] (-1,-3) -- (-3,-3) -- (-3,-0.35);
\draw[very thick,->,line width=1mm] (-5,0) -- (-3-0.35,0);
\draw[very thick,->,line width=1mm] (5,-3) -- (3+0.35,-3);

\draw[very thick] (-5,-0.5) node {\Huge $\omega$};
\draw[very thick] (5,-3+0.5) node {\Huge $d$};
\draw[very thick] (-2,0.5) node {\Huge $u$};
\draw[very thick] (2,0.5) node {\Huge $y$};
\draw[very thick] (-2,0.5-3) node {\Huge $r$};
\draw[very thick] (2,0.5-3) node {\Huge $v$};

\draw[very thick] (-3,0) node {\Large $+$};
\draw[very thick] (3,-3) node {\Large $+$};

\end{tikzpicture}

%% file: non-asymptotic-dynamic/examples-v2.tex
\begin{figure}[htb!]
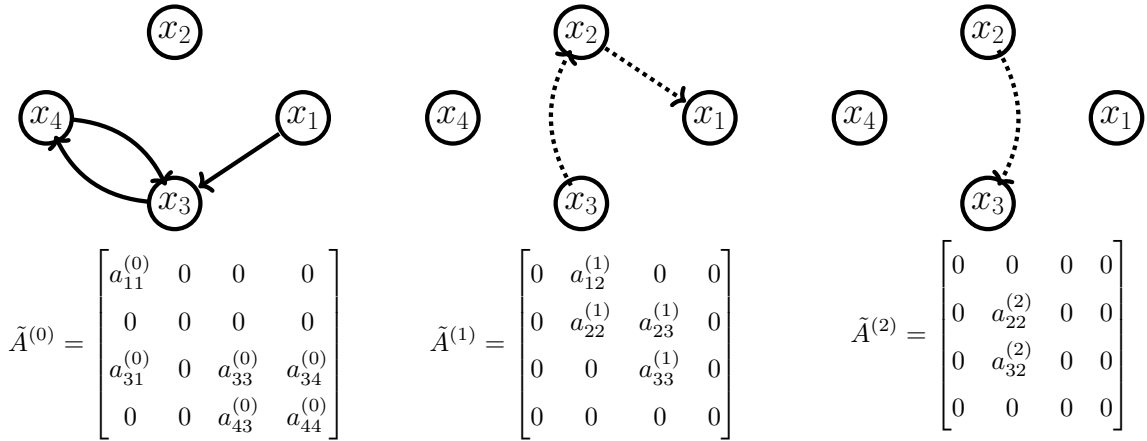

\begin{center}
\begin{minipage}[c]{.32\textwidth}
\begin{center}
\includestandalone[width=0.8\linewidth]{./fig/Azero}
\end{center}
\end{minipage}
\begin{minipage}[c]{.32\textwidth}
\begin{center}
\includestandalone[width=0.8\linewidth]{./fig/Aone}
\end{center}
\end{minipage}
\begin{minipage}[c]{.32\textwidth}
\begin{center}
\includestandalone[width=0.8\linewidth]{./fig/Atwo}
\end{center}
\end{minipage}
\begin{minipage}[c]{.32\textwidth}
\begin{equation*}
\tilde{A}^{(0)} = \begin{bmatrix}
a^{(0)}_{11} & 0 & 0 & 0 \\
0 & 0 & 0 & 0 \\
a^{(0)}_{31} & 0 &  a^{(0)}_{33} & a^{(0)}_{34} \\
0 & 0 & a^{(0)}_{43} & a^{(0)}_{44}
\end{bmatrix} 
\end{equation*}
\end{minipage}
\begin{minipage}[c]{.32\textwidth}
\begin{equation*}
\tilde{A}^{(1)}  = 
\begin{bmatrix}
0 & a^{(1)}_{12} & 0 & 0 \\
0 & a^{(1)}_{22}  & a^{(1)}_{23} & 0 \\
0 & 0 & a^{(1)}_{33} & 0 \\
0 & 0 & 0 & 0
\end{bmatrix}
\end{equation*}
\end{minipage}
\begin{minipage}[c]{.32\textwidth}
\begin{equation*}
\tilde{A}^{(2)}  = 
\begin{bmatrix}
0 & 0 & 0 & 0 \\
0 &  a^{(2)}_{22} & 0 & 0 \\
0 & a^{(2)}_{32} & 0 & 0 \\
0 & 0 & 0 & 0
\end{bmatrix}
\end{equation*}
\end{minipage}
\end{center}
\caption{\sf Illustration of network dynamics in \eqref{eq:dynamics-original}. Reliable links are shown in solid lines, and unreliable links are shown in dashed lines.}
\label{fig:stoch-uncertainty-illustration}
\end{figure}

In the setting of Section~\ref{sec:stochastic-stability-new}, $\{\triangle_j\}_{j \in [m]}$ are interpreted as multiplicative stochastic uncertainties. In this context, \eqref{eq:dynamics-mult-uncertainty} allows to model network dynamics in the presence of unreliable links as follows. Let $x(t+1)=\tilde{A}^{(0)}x(t) + \omega(t)$ be the dynamics induced by reliable links. 
Let there be $m$ links whose active status at $t$ is given by binary variables $\{\gamma_j(t)\}_{j \in [m]}$, i.e., $\gamma_j(t)=1$ if link $j$ is active at time $t$ and $\gamma_j(t)=0$ otherwise. Let the unreliable links be associated with $\{\tilde{A}^{(j)}\}_{j \in [m]}$ (see Figure~\ref{fig:stoch-uncertainty-illustration} for an illustration) such that the dynamics in the presence of reliable and unreliable links is
\begin{equation}
\label{eq:dynamics-original}
x(t+1)=\left(\tilde{A}^{(0)} + \sum_{j \in [m]} \gamma_j(t) \tilde{A}^{(j)}\right)x(t) + \omega(t)
\end{equation}

Let $\Gamma(t)=[\gamma_1(t), \ldots, \gamma_m(t)]^T$ be modeled as a multivariate Bernoulli random process which is temporally independent, and has an identical distribution at all times. Let 
$\E[\Gamma(t)] \equiv \mu = [\mu_1, \ldots, \mu_m]^T$, and let $\Sigma_{\Gamma}$ denote the constant covariance. Let
$\triangle_j(t) := \frac{\gamma_j(t)-\mu_j}{\sqrt{\Sigma_{\Gamma,jj}}}, j \in [m]$. 
Therefore, $\E[\triangle]=\E[(\triangle_1, \ldots, \triangle_m)^T]=\zerobf$, $\Sigma_{\triangle,jj}=1$ for all $j \in [m]$, and \eqref{eq:dynamics-original} can be rewritten as
$$
x(t+1)=\Big(\underbrace{\tilde{A}^{(0)} + \sum_{j \in [m]} \mu_j \tilde{A}^{(j)}}_{\Azero} + \sum_{j \in [m]} \triangle_j(t) \underbrace{\sqrt{\Sigma_{\Gamma,jj}} \tilde{A}^{(j)}}_{A^{(j)}}\Big)x(t) + \omega(t)
$$
%
which is now in the form of \eqref{eq:dynamics-mult-uncertainty}.

The above formulation has been used to study robustness of distributed averaging dynamics to link failures in 
\cite{Wang.Elia:12}, in the special case of diagonal $\Sigma_{\Gamma}$ and diagonal $\Sigma_{\omega}$, as follows. $\tilde{A}^{(0)}=I$. For every $j \in [m]$, $\tilde{A}^{(j)}$ is set as follows. Initialize $\tilde{A}^{(j)}=\zerobf$; for every link $(i,k)$ associated with $j$, add $-e_i$ to the $i$-th column and $e_i$ to the $k$-th column of $\tilde{A}^{(j)}$. Note that, in this case, all rows sums of $\Azero$ are equal to one, and, indeed $\rho(\Azero)=1$. While Theorem~\ref{thm:stable-main} does not apply directly, \cite{Wang.Elia:12} provides a  decomposition of the state into the \emph{conserved} state and the \emph{deviation} state. The dynamics of the deviation state is decoupled from that of the conserved state, and moreover the spectral radius of its nominal dynamics is less than 1, and therefore amenable to Theorem~\ref{thm:stable-main}.

\paragraph{Future Research Directions}
We remarked at the beginning of Section~\ref{sec:applications} that the $M-\triangle$ framework can be used for tight robustness analysis of linear network dynamics in several deterministic settings. It would be interesting to give a network theoretic interpretation for the small gain condition in these settings. Similarly, network-theoretic interpretations for the loop gain operator and its Perron Frobenius eigen-matrix, which play an important role in characterizing stochastic stability of linear network dynamics with multiplicative stochastic disturbances, would be of interest. The sufficient condition on $\bigtriangle$ under which Theorem~\ref{thm:stable-main} holds true is to ensure ``white-like" property for the signals in the feedback loop. Such a property allows to relate covariance of input and output signals across blocks, analogous to \eqref{eq:input-output-stable}, and hence facilitate small gain type analysis. 
Extending the analysis to $\bigtriangle$ which is correlated in time, as also noted in \cite{Bamieh.Filo:18}, would be a good step towards achieving generality in $\bigtriangle$ that is possible in the deterministic setting.

%% file: fig/Azero.tex
\begin{tikzpicture}
\draw[very thick,line width=1mm] (3,0) circle(0.6) node (a) {\Huge $x_1$};
\draw[very thick,line width=1mm] (0,2) circle(0.6) node (b) {\Huge $x_2$};
\draw[very thick,line width=1mm] (0,-2) circle(0.6) node (c) {\Huge $x_3$};
\draw[very thick,line width=1mm] (-3,0) circle(0.6) node (d) {\Huge $x_4$};

\draw[very thick,->,line width=1mm] (a) -- (c);
\draw[very thick,->,line width=1mm] (c) to[bend left] (d);
\draw[very thick,->,line width=1mm] (d) to[bend left] (c);

\end{tikzpicture}

%% file: fig/Aone.tex
\begin{tikzpicture}
\draw[very thick,line width=1mm] (3,0) circle(0.6) node (a) {\Huge $x_1$};
\draw[very thick,line width=1mm] (0,2) circle(0.6) node (b) {\Huge $x_2$};
\draw[very thick,line width=1mm] (0,-2) circle(0.6) node (c) {\Huge $x_3$};
\draw[very thick,line width=1mm] (-3,0) circle(0.6) node (d) {\Huge $x_4$};

\draw[very thick,dashed,->,line width=1mm] (b) -- (a);
\draw[very thick,dashed,->,line width=1mm] (c) to[bend left] (b);

\end{tikzpicture}

%% file: fig/Atwo.tex
\begin{tikzpicture}
\draw[very thick,line width=1mm] (3,0) circle(0.6) node (a) {\Huge $x_1$};
\draw[very thick,line width=1mm] (0,2) circle(0.6) node (b) {\Huge $x_2$};
\draw[very thick,line width=1mm] (0,-2) circle(0.6) node (c) {\Huge $x_3$};
\draw[very thick,line width=1mm] (-3,0) circle(0.6) node (d) {\Huge $x_4$};

\draw[very thick,dashed,->,line width=1mm] (b) to[bend left] (c);

\end{tikzpicture}

%% file: capacity-dynamic/network-flow-dynamics-v2.tex
\section{Robustness of Finite Networks: Nonlinear Dynamical Setting}
\label{sec:capacity-perturbation-dynamic}

\input{./capacity-dynamic/pls}
\subsection{Robustness under Information Constraint}
\label{sec:routing} 
We discuss two setups to illustrate different ways in which information constraint on the controller can affect network robustness. In the setup in Section~\ref{subsubsec:loss}, information constraint results in loss in robustness, where there is no such loss in the setup in Section~\ref{subsubsec:no-loss}. 

\subsubsection{Loss in Robustness under Decentralized Control}
\label{subsubsec:loss}
We return to the setup of \eqref{eq:network-flow-dynamics-special-1}. We let $f_i(x) \equiv f_i(x_i)$ be strictly increasing and have capacity constraint $f_i(x_i) \leq c_i$, for all $i \in \mc E$. We simplify by letting $R_{ji}(x)\equiv R_{ki}(x)\equiv R_i(x)$ for any two links $j$ and $k$ incident on to $i$, i.e., each node adds flow from immediate upstream links and instantaneously routes it among its immediate downstream links, without consideration for the source or destination of that flow.
%
This naturally makes sense for a \emph{single commodity} flow. For simplicity in presentation, we assume single source and single sink. To avoid triviality, we assume that,  
for every node $v \in \mc V$, there exists a directed path from the source node to $v$ and from $v$ to the sink node.
Extension to multiple sources and sinks, for single commodity flow, follows from the construction in Section~\ref{sec:notations}. Correspondingly, we shall remove subscript and let $\lambda$ denote the external flow to the only source node. 

In the current setting, the stability and robustness properties of the network flow dynamics depend on the choice of routing functions $R$. Of particular interest is decentralized routing where the routing control action for a link depends only on the state on $i$ and on links having the same tail node as $i$, i.e., $R_i(x) \equiv R_i(\{x_j: \, j \text{ and } i \text{ have same tail node}\})$. Consider a feasible flow $f^*$, i.e., satisfying the flow conservation and capacity constraints, and correspondingly $\{x_i^*:=f_i^{-1}(f_i^*)\}_{i \in \mc E}$. Given such $x^*$ and $f^*$, which exist if and only if $\lambda$ is less than the minimum cut capacity of the network, consider decentralized routing control satisfying: (i) $R_i(x^*)=\frac{f_i^*}{\sum_{j}f_j^*}$ for all $i \in \mc E$, where the summation is over all links $j$ which have the same tail node as $i$, and (ii) $\frac{\partial R_i(x)}{\partial x_j} >0$ for every two distinct links $i$ and $j$ having the same tail node. (i) implies that $x^*$ is an equilibrium. (ii) implies that, if $x_j$ increases but the state on other links which affect $R_i$ remain the same, then more flow will be routed to every other link. This property helps to establish global asymptotic stability of $x^*$ if $\mc G$ is acyclic. This is because (ii) along with increasing nature of $f_i(.)$ implies that the Jacobian of \eqref{eq:network-flow-dynamics-general} under these conditions is a compartmental matrix, i.e., a matrix whose off-diagonal components are non-negative and column sums are non-positive. Such systems are known to possess a contraction principle, e.g., see \cite{Sontag:10}. 

The (decentralized) routing satisfying (i) and, particularly, (ii) above is referred to as \emph{monotone}  routing. An example of monotone routing is:
\begin{equation*}
R_i(x) = \frac{f_i^* e^{\beta(x^*_i-x_i)}}{\sum_j f_j^* \exp(\beta(x^*_j-x_j))}, \quad \beta>0
\end{equation*}
which is the multinomial logit model at node $v$ for discrete choice among links in $\mc E_v^+$, when the utility function associated with $i \in \mc E_v^+$ is $x_i^*-x_i+\log f_i^*$. Indeed, monotone routing is interpreted as en route driver decision in \cite{Como.Savla.ea:SICON11}, with $x^*$ being the nominal network state expected by the drivers during the trip. This nominal state is updated by the driver population at a slower time scale according to a best response rule, e.g., see \cite{Sandholm:10}. The link (out-)flows under such a multi-scale update rule are shown to converge to a Wardrop equilibrium in \cite{Como.Savla.ea:SICON11}, if $\mc G$ is acyclic, and if each $f_i(.)$, $i \in \mc E$, is strictly concave in addition to being strictly increasing. 

The monotone routing policies are also shown to admit the following sharp characterization of robustness of equilibrium $x^*$ to perturbation in capacity~\cite{Como.Savla.ea:Part1TAC10,Como.Savla.ea:Part2TAC10}. Let $\triangle \in [0,c]$ denote the perturbation in capacity. Clearly, there exists a $\|\triangle\|_1$ (infinitesimally) greater than the network residual capacity, such that at least one component of $x(t)$ grows unbounded for every (not necessarily decentralized) routing policy. Such a $\triangle$, e.g., corresponds to reducing capacities on the links outgoing from the minimum cut by infinitesimally greater than the residual capacities on them. However, if the routing policy is decentralized, such as the monotone routing, then reducing capacities on the links, outgoing from the node having the smallest residual capacity, by an amount infinitesimally greater than their respective residual capacities is sufficient to ensure that $x_i(t)$ grows unbounded at least for one $i \in \mc E$. 
Interestingly, under the monotone routing, if $\mc G$ is acyclic, then for \emph{every} $\|\triangle\|_1$ less than the minimum node residual capacity, there exists a globally asymptotically stable equilibrium, which is different than the equilibrium $x^*$ associated with the nominal, i.e., non-perturbed system. The existence of a new equilibrium is established by a novel use of the monotonicity property of the underlying dynamics in \cite{Como.Savla.ea:Part2TAC10}, and the stability then follows from the contraction principle, along the same lines as the stability of nominal dynamics. 

In summary, if one interprets the minimum node residual capacity to be the robustness under decentralized routing, then this is in general strictly less than the network residual capacity, which is the robustness under a centralized routing policy. Section~\ref{sec:routing+scheduling} discusses how allowing additional control action in this setting can prevent such a loss. We next discuss a different setup where decentralized control does not cause loss in robustness.

\subsubsection{No Loss in Robustness under Decentralized Control}
\label{subsubsec:no-loss}
Let us fix the routing matrix; therefore the decoupling of $\interflow_{ij}$ from Section~\ref{subsubsec:loss} becomes $\interflow_{ij}(x)=f_i(x) R_{ij}$. This allows us to consider multiple commodity flow in between multiple sources and sinks. 
%
We remove $\linkin$ and let the external flow arrive directly on to the links outgoing from the source nodes, similar to \eqref{eq:network-flow-linear}. In summary, in the current setting, \eqref{eq:network-flow-dynamics-general} becomes:
\begin{equation}
\label{eq:network-flow-dynamics-scheduling}
\dot{x} = \left(R^T - I \right) f(x) + \lambda
\end{equation}
where $\lambda \in \reals_{\geq 0}^{\mc E}$ is the external inflows at the links, and, with a slight abuse of notation, $R$ denotes the routing sub-matrix restricted to links in $\mc E$. 
To avoid triviality, we assume that for every node $v \in \mc V$, there exists a directed path from at least one source node to $v$, and there exists a directed path from $v$ to at least one sink node.

Motivated by scenarios involving allocation of a fixed service resource to multiple conflicting queues, e.g., at a signalized traffic intersection in urban traffic network or at a router in communication network, consider the following simple model for link outflows:
\begin{equation}
\label{eq:outflow-scheduling}
f_i(x) \equiv f_i(x_i)= \Big \{\begin{array}{ll}
u_i(x) c_i & \text{ if }x_i > 0 \\
\sum_{j} R_{ji} f_j(x) & \text{ if }x_i=0
\end{array}, \qquad i \in \mc E
\end{equation}
where the summation is over all links $j \in \mc E \cup \linkin$ that are incident to $i$. The control input $u_i(x) \in [0,1]$ can be interpreted as the fraction of times that the queue on $i$ gets serviced. \eqref{eq:outflow-scheduling} ensures positivity of \eqref{eq:network-flow-dynamics-scheduling} in the current setting. The fixed resource feature is modeled by imposing 
\begin{equation}
\label{eq:finite-resource}
\sum_{i \in \mc E_v^-} u_i(x) \leq 1,  \quad  v \in \mc V
\end{equation}
By further imposing $u_i(x) \to 0^+$ as $x_i \to 0^+$ (and starting from an initial condition $x_0 > \zerobf$), one can ensure that the $x_i=0$ case in \eqref{eq:outflow-scheduling} is never active, and thereby also avoiding regularity issues with the right hand side of \eqref{eq:network-flow-dynamics-scheduling}.

A necessary condition for the boundedness of $x(t)$, and hence also for the existence of equilibrium, in the current setting, is $\sum_{i \in \mc E_v^-} \frac{f_i^*}{c_i} \leq 1$ for all $v \in \mc V$, where $f^*$ is the flow \emph{induced} by $\lambda$ and $R$, i.e., $\lambda+(R^T-I)f^*=0$, which implies $f^*=(I-R^T)^{-1} \lambda$. It is of interest to study stability conditions under natural controllers $u(x)$ and compare with the necessary condition. It is straightforward to see that, if the necessary condition is satisfied, then the open-loop controller $u_i(x) \equiv f_i^*/c_i$, $i \in \mc E$, satisfies \eqref{eq:finite-resource}, and, under it,  every $x>\zerobf$ is an equilibrium of \eqref{eq:network-flow-dynamics-scheduling}. Besides being open-loop, this controller requires information about $\lambda$ and $R$ and is therefore centralized. Similar to Section~\ref{subsubsec:loss}, it is of interest to study decentralized controllers of the form: $u_i(x) \equiv u_i(x_j: \, j \text{ has the same head node as }i)$. Using monotonicity arguments similar to Section~\ref{subsubsec:loss}, it is shown in \cite{Savla.Lovisari.ea.Allerton13} that, if $\mc G$ is acyclic and if the necessary condition is satisfied, then, under the following controller, there exists a unique $x^* > \zerobf$ whose basin of attraction is $\reals_{>0}^{\mc E}$:  
\begin{equation}
\label{eq:pf-controller}
u_i(x) = \frac{x_i}{\sum_{j} x_j + \beta}, \qquad \beta>0
\end{equation}
where the summation is over all $j$ with the same head node as $i$. Extension of this result to cyclic $\mc G$ is possible by establishing $\sum_{i \in \mc E} x_i \log \left( u_i(x) \frac{c_i}{f_i^*}\right)$ as Lyapunov function. The inspiration for this analysis comes from the literature on stochastic networks, e.g., see \cite{Massoulie:07}, where \eqref{eq:pf-controller} is interpreted as a \emph{proportionally fair} controller for the current setting. It is remarkable to note that the decentralized controller in \eqref{eq:pf-controller} gives the same stability guarantee as the centralized (open-loop) controller without requiring information about (even local) $\lambda$, $R$ or $c$. The literature on back pressure controllers, e.g., see \cite{Tassiulas.Ephremides:92}, suggests that one can get the same stability guarantee by establishing $\sum_{i \in \mc E} x_i^2$ as Lyapunov function. 
 Such tight stability guarantees translate into tight guarantees on robustness to perturbation to link capacities. That is, for a perturbation in capacity, if there exists a centralized controller under which $x(t)$ remains bounded, then it does so also under the above decentralized controllers. Hence, no loss in robustness under decentralized control.  

\subsection{Compensating Information Constraint with Control Action for Robustness} 
\label{sec:routing+scheduling}
Consider again the single commodity flow setting, with a single source and a single sink from Section~\ref{subsubsec:loss}. 
Recall that, in Section~\ref{subsubsec:loss}, the outflow from a link was uncontrolled and determined by $\{f_i(x_i)\}_{i \in \mc E}$. Let us relax this feature, but maintain the decentralization aspect. 
That is, we let $\interflow_{ij}(x)$ be controlled, and let $\interflow_{ij}(x) \equiv \interflow_{ij}(\{x_k: \, k=i \text{ or $k$ has the same tail node as }j\})$. An implication of the relaxation is that  we extend the scope of control to also \emph{schedule} the outflow from a link in addition to determining the routing part. A natural extension of the monotonicity property from Section~\ref{subsubsec:loss} to the current setting is to consider decentralized $\{\interflow_{ij}\}$ which satisfy the following for all $i \in \mc E \cup \linkin$ and $j \in \mc E \cup \linkout$: (i)  $\frac{\partial \interflow_{ij}(x)}{\partial x_k} >0$ for $k=i$, and for all $k$, except $j$, which has the same tail node as $j$, and (ii) $\frac{\partial \sum_{j} \interflow_{ij}(x)}{\partial x_k} <0$ for all $k$ to which $i$ is incident. These properties imply that if $x_k$ increases, but the state on the other links which affect $\interflow_{ik}$ remains the same, then (i) more flow is sent to links, other than $k$, but which have the same tail node as $k$, and (ii) the total outflow $\sum_{j} \interflow_{ij}(x)$ from every link $i$ incident on to $k$ decreases. We call the control policies $\{\interflow_{ij}(x)\}$ satisfying these two properties as \emph{monotone control policies} (in contrast to monotone \emph{routing} policies in Section~\ref{subsubsec:loss}) with a slight abuse of terminology. An example of monotone control is:
$$
\interflow_{ij}(x) = c_i \left(1 - e^{- \beta x_i} \right) \frac{e^{- \beta x_j}}{\sum_k e^{-\beta x_k}}, \qquad \beta>0
$$
where the summation is over $k=i$ and all $k$ which have the same tail node as $j$. 

In \cite{Como.Lovisari.ea:TCNS13}, using similar tools as described in Section~\ref{subsubsec:loss}, it is shown that, under monotone control, if $\lambda$ is less than the min cut capacity, then there exists a globally asymptotically stable equilibrium $x^*$. An implication of this is that, for all perturbations to capacity $\|\triangle\|_1$ less than the network residual capacity, there exists a globally asymptotically stable (new) equilibrium under monotone control. Combining this with the fact that, there exists $\triangle$ with $\|\triangle\|_1$ (infinitesimally) greater than the network residual capacity such that no controller (not necessarily decentralized) can prevent $x(t)$ from growing unbounded on at least one component implies that monotone control policies are maximally robust to perturbation in capacity among all (not necessarily decentralized) controllers. This robustness guarantee is stronger than the one on monotone routing policy in Section~\ref{sec:routing} which further required acyclic assumption on $\mc G$. 
Moreover, the monotone routing is parameterized by equilibrium $x^*$ and hence requires some kind of centralized coordination, whereas no such coordination is required in the current monotone control setting. The additional control action in the current setting in the form of scheduling, in conjunction with property (ii) in the definition of monotone control, allows for forward \emph{and} backward propagation of dynamics. Interestingly, such propagations are sufficient to compensate for the decentralization feature of the controller to give as good robustness performance as any centralized controller. On the other hand, the lack of scheduling action in the setting of Section~\ref{sec:routing} allows only forward propagation of dynamics resulting in gap in robustness guarantees with respect to a centralized controller. 

\cite{Como.Lovisari.ea:TCNS13} also considers a few extensions, most notable being consideration of finite storage capacity on the links, i.e., $x(t) \leq B \in \reals_{>0}^{\mc E}$. 
If $\lambda$ is less than the minimum cut capacity (which is a necessary condition for boundedness of $x(t)$), it is shown that there exists a globally asymptotic stable equilibrium  under appropriate modification of the monotone controller. It is also shown that, under such a modified controller, if the necessary condition does not hold true, then there exists a cut such that all links outgoing from that cut hit their respective storage capacities (which are not necessarily equal) simultaneously. This possibly suggests \emph{graceful degradation} in the unstable regime, i.e., maximizing the time till any link hits storage capacity which could cause link failure. 

%

 

\paragraph{Future Research Directions} 
The discussion in the beginning of this section suggests lack of understanding of the connection between analysis techniques for piecewise linear systems, which potentially are computationally intense and conservative, and the nonlinear analysis techniques which are proving to be effective in specific instances of nonlinear network dynamics. Comparing the techniques in these specific cases could be a good first step towards understanding the connection. It is expected to be challenging however to treat structural constraints such as distributed information and control actions in a similar fashion. Developing specialized tools to analyze the impact of such constraints on network robustness could be of independent interest. Some specific future directions along these lines are robustness analysis under general, e.g., multi-hop information constraint in the setting of Section~\ref{subsubsec:loss} and generalizing the substitutability between control actions and global information beyond the setup in Section~\ref{sec:routing+scheduling}.

%% file: capacity-dynamic/pls.tex
%
An interesting class of network dynamics is network flow dynamics. To formulate it, recall the notions of super source and super sink from Section~\ref{sec:notations}. It is convenient to identify them both with the same virtual node, say $0$. Add a virtual directed link $(0,v)$ from the virtual node to every source node $v \in \mc V$; let this set be denoted as $\linkin$. Similarly add a virtual directed link $(v,0)$ from every sink node $v \in \mc V$ to the virtual node; let this set be denoted as $\linkout$. Network flow dynamics corresponds to mass conservation:
\begin{equation}
\label{eq:network-flow-dynamics-general}
\dot{x}_i = \underbrace{\sum_{j \in \mc E \cup \linkin} \interflow_{ji}(x)}_{\text{inflow}} - \underbrace{\sum_{j \in \mc E \cup \linkout} \interflow_{ij}(x)}_{\text{outflow}}, \qquad i \in \mc E
\end{equation}
%
 where $x_i \geq 0$ denotes the mass on link $i \in \mc E$, and $\interflow_{ij} \geq 0$ is the flow from link $i \in \mc E \cup \linkin$ to link $j \in \mc E \cup \linkout$. Naturally, $\interflow_{ij}=0$ if $i$ is not incident to $j$. The $\interflow_{ij}$'s depend on the state $x$, where this dependence also incorporates feedback control; the exact model depends on the setting. In all these settings, $\interflow_{ij}$'s are constrained to be such that $\interflow_{ij}(x)=0$ for all $j$ if $x_i=0$,  and $\sum_{j \in \mc E \cup \linkout} \interflow_{ij}(x) = \lambda_i$ if $i \in \linkin$. The first constraint ensures that \eqref{eq:network-flow-dynamics-general} is a positive system, i.e., $x(t) \geq \zerobf$ for all $t$, and the second constraint specifies the total external flow coming into source nodes.

It is natural to consider the decomposition $\interflow_{ij}(x)=f_i(x) R_{ij}(x)$, where $f_i(x) \geq 0$ is the net outflow from $i \in \mc E \cup \linkin$ satisfying $f_i(x)=0$ if $x_i=0$ for all $i$, and $f_i(x) \equiv \lambda_i$ for all $i \in \linkin$, and $R_{ij}(x) \in [0,1]$ is the routing function satisfying $R_{ij}(x) \equiv 0$ if $i$ is not incident to $j$, and $\sum_{j} R_{ij}(x) \equiv 1$ for all $i$. Under these assumptions, \eqref{eq:network-flow-dynamics-general} can be written in vector form as:
\begin{equation}
\label{eq:network-flow-dynamics-special-1}
\dot{x}=\left(R^T(x) - I \right) f(x) + \lambda
\end{equation}
A simple linear version of \eqref{eq:network-flow-dynamics-special-1} is obtained by assuming constant routing matrix, i.e., $R(x) \equiv R$, and let the outflow from $i$ be linearly increasing in $x_i$, i.e., $f_(x) \equiv f_i(x_i) = h_i x_i$ for $h_i>0$. Let $H$ be the diagonal matrix whose entries are $\{h_i\}_{i \in \mc E}$. Since $R$ is fixed, we can let the external flow arrive directly on to the links outgoing from the source nodes. Therefore, it is sufficient to consider \eqref{eq:network-flow-dynamics-special-1} restricted to links in $\mc E$ as:
\begin{equation}
\label{eq:network-flow-linear}
\dot{x} = \left(R^T - I \right) H x + \lambda
\end{equation}
where, we use the same notation as \eqref{eq:network-flow-dynamics-special-1} for brevity. 
Invertibility of $R^T-I$ follows from the connectivity of the underlying graph, and the 
fact that $R$ (restricted to $\mc E$) is row sub-stochastic with the entries of at least one row, corresponding to a sink, adding to strictly less than 1. Therefore, \eqref{eq:network-flow-linear} admits a unique equilibrium $x^*=H^{-1}(I-R^T)^{-1} \lambda$, whose stability analysis is straightforward. 

A key practical consideration for network flow dynamics is to include link-wise capacity constraints by saturating link-wise outflows at $\{c_i\}_{i \in \mc E}$ (cf. ``Network Flow" in Section~\ref{sec:notations}). Introducing such a saturation in \eqref{eq:network-flow-linear} leads to piecewise affine system, which can be written in the state-space form as:
\begin{equation}
\label{eq:network-flow-piecewise-linear}
\begin{split}
\dot{x} & = \underbrace{\left(R^T - I \right) H}_{A} x + \underbrace{\left(I - R^T \right)}_{B} u + \lambda \\
y & = \underbrace{H}_{C} x \\
u(t) & = \max \left\{0, y(t) - c\right\}
\end{split}
\end{equation}
\begin{figure}
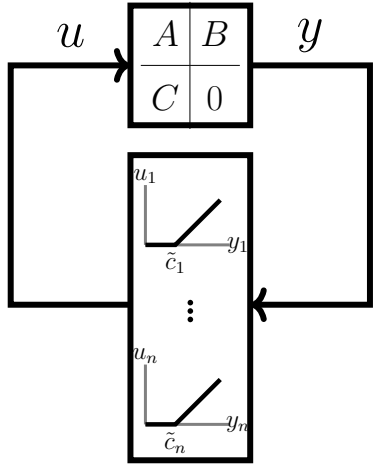

\begin{center}
\includestandalone[width=0.3\textwidth]{./fig/networkflow-pls}
\caption{An LTI system in feedback with on-off nonlinearities.}
\label{fig:networkflow-PLS}
\end{center}
\end{figure}
where $c$ is the vector of link-wise capacities and $\max$ is element-wise. \eqref{eq:network-flow-piecewise-linear}, which can be converted to a piecewise \emph{linear} system (PLS) through a simple change of variable along with corresponding change of $c$ to say $\tilde{c}$, can be interpreted as an LTI system in feedback with on-off nonlinearities; see Figure~\ref{fig:networkflow-PLS}. Sufficient condition for global stability and robustness analysis of PLS have been developed, e.g., see \cite{Goncalves.Megretski.ea:03} and references therein. While these tools have not been applied to the specific PLS in \eqref{eq:network-flow-piecewise-linear} to the best of our knowledge, such an analysis will potentially face a few challenges. First, it remains to be seen how does the sufficient condition for global asymptotic stability compare with the necessary condition given by the max flow min cut theorem in the single sink setting. Second, the number of \emph{switching surfaces} in \eqref{eq:network-flow-piecewise-linear} grows exponentially with the number of links, and thereby making the analysis computationally intense. Third, it is not clear how the analysis can be extended to other interesting nonlinear instances of \eqref{eq:network-flow-dynamics-general}. In the remainder of this section, we present alternate nonlinear techniques for stability analysis of \eqref{eq:network-flow-dynamics-general} and robustness to perturbations in $\{c_i\}_{i \in \mc E}$,  under different setups for $\{\interflow_{ij}(x)\} \in \interflowconstraints$. In Section~\ref{sec:routing}, we discuss how information constraint with respect to centralized control reduces robustness, and in Section~\ref{sec:routing+scheduling}, we discuss a setup where additional control action compensates for information constraint to maintain the same robustness as under centralized control.  

%% file: fig/networkflow-pls.tex
\begin{tikzpicture}
\draw[very thick,line width=1mm] (-1,-1) rectangle (1,1);
\draw[very thick] (0,0) node {\LARGE $\begin{array}{c|c}
A & B \\
\hline
C & 0
\end{array}$};

\draw[very thick,line width=1mm] (-1,-6.6) rectangle (1,-1.5);
\draw[very thick] (0,-4) node {\Huge $\vdots$};

\draw[very thick,->,line width=1mm] (1,0) -- (3,0) -- (3,-4) -- (1,-4);

\draw[very thick,->,line width=1mm] (-1,-4) -- (-3,-4) -- (-3,0) -- (-1,0);

\draw[very thick] (-2,0.5) node {\Huge $u$};
\draw[very thick] (2,0.5) node {\Huge $y$};

\draw[gray, line width=0.5mm] (-0.75,-3) -- (-0.75,-2);
\draw[gray, line width=0.5mm] (-0.75,-3) -- (0.65,-3);
\draw[very thick] (-0.75,-1.85) node {\large $u_1$};
\draw[very thick] (0.8,-3) node {\large $y_1$};
\draw[very thick, line width=0.75mm] (-0.75,-3) -- (-0.25,-3) -- (0.5,-2.25);
\draw[very thick] (-0.25,-3.3) node {\large $\tilde{c}_1$};

\draw[gray, line width=0.5mm] (-0.75,-6) -- (-0.75,-5);
\draw[gray, line width=0.5mm] (-0.75,-6) -- (0.65,-6);
\draw[very thick] (-0.75,-4.85) node {\large $u_n$};
\draw[very thick] (0.8,-6) node {\large $y_n$};
\draw[very thick, line width=0.75mm] (-0.75,-6) -- (-0.25,-6) -- (0.5,-5.25);
\draw[very thick] (-0.25,-6.3) node {\large $\tilde{c}_n$};

\end{tikzpicture}

%% file: cascading-failure/cascading-failure-main-v2.tex
\section{Robustness under Cascading Failure}
\label{sec:cascading-failure}

\input{./cascading-failure/cascading-failure-network-flow}

\input{./cascading-failure/cascading-failure-contagion}

\paragraph{Future Research Directions} 
A very few formal stability and robustness analyses exist for cascading failure dynamics in physical networks. It would be natural to investigate if the nonlinearities induced by cascading failure can be cast into canonical forms, similar in spirit to the capacity constraint in Section~\ref{sec:capacity-perturbation-dynamic}, so as to use tools from the piecewise linear or hybrid systems literature.
In the setup in Sections~\ref{sec:cascade-electrical} and \ref{sec:cascade-transport}, vulnerability of a link is determined by its residual capacity with respect to the nominal equilibrium flow, which in turn could be the outcome of a process. This is to be contrasted with the setting in Section~\ref{sec:cascading-failure-contagion}, where assignment of thresholds to nodes is independent of the initial network structure or the network formation process which leads to the initial interconnection structure. Modeling this dependency and  evaluating resilience of the network formation process will be an interesting direction to pursue.

%% file: cascading-failure/cascading-failure-network-flow.tex
The notion of capacity was introduced in Sections~\ref{sec:capacity-perturbation-static} and \ref{sec:capacity-perturbation-dynamic} by \emph{thresholding} the link (out-)flows. The capacities potentially define boundary to failure of the corresponding link. The failure upon crossing the boundary implies a structural change in $\mc G$. The complete description then necessitates specifying the jump map under such a structural change, i.e., the network flow dynamics immediately after the discontinuity induced by the failure. The jump could further result in crossing of the capacity for some other link, potentially leading to a series of failures -- a phenomenon known as \emph{cascading failure}. Notationally, this is denoted as $\{\mc G(t)=(\mc V, \mc E(t), \mc W(t)): \, t = 0, 1, 2, \ldots\}$, with $\mc E(0) \supset \mc E(1) \supset \ldots$, where $\mc G(0)$ is the nominal graph structure. 

Given the finite size of $\mc G(0)$, the process of cascading failure terminates after finite steps, either at a subnetwork with link-wise flows less than the respective link-wise capacities, or at a subnetwork in which there is no directed path from the source node to the sink node. The former scenario corresponds to  a new equilibrium for the network structure, and the latter corresponds to network failure. It is of interest to tightly characterize the set of perturbations under which the cascading failure terminates at an (new) equilibrium network structure.    
In Sections~\ref{sec:cascade-electrical} and \ref{sec:cascade-transport}, we consider jumps governed by control and physical constraints on flow, along the lines of Sections~\ref{sec:capacity-perturbation-static} and \ref{sec:capacity-perturbation-dynamic}. 
We then discuss a representative contagion model for cascading failure, and present brief remarks at the end of the section on its contrast with respect to the models in Sections~\ref{sec:cascade-electrical} and \ref{sec:cascade-transport}.


\input{./cascading-failure/electrical-networks}

\input{./cascading-failure/transport-networks}

%% file: cascading-failure/electrical-networks.tex
\subsection{Electrical Networks}
\label{sec:cascade-electrical}
Consider the setup of electrical networks from Section~\ref{sec:capacity-perturbation-static} with fixed link weights $\mc W$. Failure of a link, say $i$, corresponds to setting $\mc W_i=0$. The network flow after such a jump is again given by \eqref{eq:electrical-flow} by setting the entries of $\mc W$, corresponding to the links which have failed, to be zero. 
A necessary and sufficient condition for cascading failure to terminate at $t$ is $\lambda<\lambdamax(\mc W(t)) =: \lambdamax(\mc W(t))$ (refer to Section~\ref{sec:capacity-perturbation-static} for the definition of $\lambdamax(\mc W)$). 
%
%
Therefore, if $\lambdamax(\mc  W(1)) \geq \lambdamax(\mc W(2)) \geq \ldots$, then a $\triangle$ causes network failure if and only if $\lambda > \lambdamax(\mc W(1))$. 
However, $\{\lambdamax(\mc W(t))\}_{t \in \naturals}$ is not monotonically decreasing in general. Therefore, in general, a $\triangle$ causes network failure if and only if $\lambda > \max_{t \in \naturals} \lambdamax(\mc W(t))$. Due to the multi-stage nature, giving a tight characterization of the set of all $\triangle$ leading to network failure in the general case is computationally challenging. 

Let us extend the robustness analysis to the case where the above jump map is also influenced by control actions. Let us consider $\mc W$ as the control action, as in Section~\ref{sec:capacity-perturbation-static}, and let $\mc W^* \in \argmax_{\mc W(0) \in [\Wmin,\Wmax]} \lambdamax(\mc W(0))$ (a problem discussed in Section~\ref{sec:capacity-perturbation-static}). If $\Wmin=\zerobf$, then, with $\mc W(0) = \mc W^*$:
$$
\lambdamax(\mc W^*) \geq  \lambdamax(\mc W(1)) \geq \lambdamax(\mc W(2)) \geq \ldots
$$
where the inequalities follow from the fact that $\Wmin=\zerobf$ implies that the choice of $\mc W^*$ includes optimizing over all possible sub-networks of $\mc G(0)$ (recall that setting $\mc W_i=0$ is equivalen to removal of link $i$). The above monotonically decreasing relationship implies not only that $\mc W^*$ is maximally robust, but also it is straightforward to quantify its robustness tightly. Formally, a $\triangle$ which removes the residual capacity $c_i - |f_i(\mc W^*,\Lambda)|$ from the link $i$ for which this value is minimum, and does not perturb other links, has the smallest 
$\|\triangle\|_1$ among all $\triangle$ which cause network failure. On the other hand, for all $\|\triangle\|_1 < \min_{i \in \mc E} \, c_i - |f_i(\mc W^*,\Lambda)|$, there is no link failure, and hence no network failure. 

\begin{figure}[htb!]
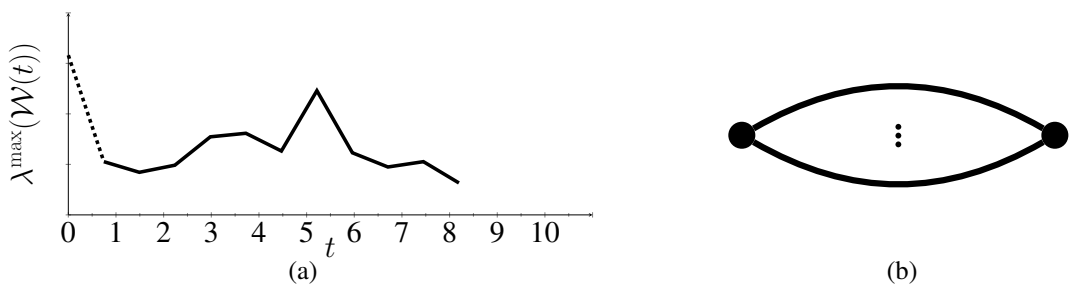

\begin{center}
\begin{minipage}[c]{.6\textwidth}
\begin{center}
\includestandalone[width=0.8\linewidth]{./fig/non-monotonicity} 
\end{center}
\end{minipage}
\begin{minipage}[c]{.35\textwidth}
\begin{center}
\includestandalone[width=0.8\linewidth]{./fig/parallel-net} 
\end{center}
\end{minipage}
\begin{minipage}[c]{.6\textwidth}
\begin{center}
(a)
\end{center}
\end{minipage}
\begin{minipage}[c]{.35\textwidth}
\begin{center}
(b)
\end{center}
\end{minipage}
\end{center}
\caption{(a) Non-monotonicity of $\lambdamax(\mc W(t))$ of an electrical network during the cascading failure process; (b) A two-node network.}
\label{fig:non-monotonicity}
\end{figure}

An alternate setting involves controlling $\lambda$ and keeping $\mc W$ fixed.  
%
In order to motivate this, let us revisit the general scenario in the uncontrolled case, illustrated in Figure~\ref{fig:non-monotonicity}(a). Every $\triangle$ such that $\lambda_0 := \lambda > \max_{t \in \naturals} \lambdamax(\mc W(t))$ causes network failure. However, if one could choose $\lambda(k) <  \max_{t \in \naturals} \lambdamax(\mc W(t))<\lambda$ for all $k \geq 1$, then the cascading process will terminate before network failure. Since $\lambda$ is typically construed as measure of network performance, this illustrates tradeoff between performance and robustness. This tradeoff is more pronounced when there is an additional requirement to terminate the cascading process within a given horizon. For example, in Figure~\ref{fig:non-monotonicity}(a), if this horizon is $6$, then the best performance is achieved for $\lambda(t)=\lambdamax(\mc W(5))$ for all $t=1, \ldots, 6$, which is strictly less than $\max_{t \in \naturals} \lambdamax(\mc W(t))$. 

%
%

When selecting the best $\{\lambda(t)\}_t$ in the above example, we implicitly assumed that the $\{\lambdamax(\mc W(t))\}_t$ is independent of $\{\lambda(t)\}_t$. However, in general, this is true only for two-node networks (cf. Figure~\ref{fig:non-monotonicity}(b)). This is because, in general, $\lambda(k)$ affects the sequence $\{\mc W(t)\}_{t \geq k}$, and hence it affects $\{\lambdamax(\mc W(t))\}_{t \geq k}$. Therefore, computing optimal $\{\lambda(t)\}_t$ for the general case is challenging. \cite{Ba.Savla:TAC17} presents a computational approach to solve this problem. Specifically, it constructs a finite abstraction in the form of a directed acyclic graph with states $(\lambda(t),\mc W(t))$, rooted at the initial condition $(\lambda(0),\mc W(0))$, and leaf nodes being desirable termination states of the cascade process. Edges correspond to one step of the cascade process. Edge costs are zero except if the edge is incoming into a leaf node, in which case the cost is equal to the value of $\lambda$ associated with the leaf node. 
Among all the paths from the root to a leaf node, one with the least cost, is shown to give an optimal $\{\lambda(t)\}_t$. This methodology holds true for the multiple sink multiple source scenario, and is particularly suited for the finite horizon termination constraint. 



%% file: fig/non-monotonicity.tex
\begin{tikzpicture}
\begin{axis}[width=1.65\textwidth,
    height=\axisdefaultheight,
    ticklabel style = {font=\Huge},
    axis x line = bottom,
    axis y line = left,
    minor x tick num=1,
    xlabel={\Huge $t$},
    xmin=0,xmax=11,
    ylabel={\Huge $\lambdamax(\mc W(t))$},
    ymin=0,ymax=2,
    extra x tick labels={\kern1mm},
    xtick={0,1,2,...,11},
    xticklabels= {0,1,2,3,4,5,6,7,8,9,10},
    yticklabels={,,}
    ]
    \end{axis}
    \draw[line width=1mm] (1,1.5) -- (2, 1.2) to (3, 1.4) to (4, 2.2) to (5,2.3) to (6,1.8) to (7,3.5) to (8,1.75) to (9,1.35) to (10,1.5) to (11,0.9); 
    \draw[dashed,line width=1mm] (0,4.5) -- (1,1.5);
       
%
%
%
%
%

\end{tikzpicture}

%% file: fig/parallel-net.tex
\begin{tikzpicture}[node distance=5cm]
\node[draw, circle, line width=1mm, scale=1, fill=black] (1) {$ $}; 
\node[draw, circle, line width=1mm, scale=1, fill=black] (2) [right of=1] {$ $};   
  
  \path[main edge, edge label] 
(1) edge[bend left, line width=1mm] node{$ $} (2)
(1) edge[bend right, line width=1mm] node{$ $} (2);
\draw (2.5,0.1) node {\Huge $\vdots$};

\end{tikzpicture}

%% file: cascading-failure/transport-networks.tex
\subsection{Transport Networks}
\label{sec:cascade-transport}
Let us now revisit the setup of transport networks from Sections~\ref{subsubsec:loss} and \ref{sec:routing+scheduling}, where the network flow dynamics immediately after failure at $t$ is determined by the routing policy restricted to residual graph $\mc G(t)$. 
In the centralized and static setting, a policy that routes over $\mc G(t)$ according to a feasible flow 
is maximally robust, with the network residual capacity of $\mc G(0)$ being the measure of robustness. 
This is because the minimum cut capacity of a network is no less than that of any of its sub-network. Analyzing robustness of decentralized routing under hybrid dynamics obtained from the combination of flow dynamics from Sections~\ref{subsubsec:loss} or \ref{sec:routing+scheduling} over a given network with dynamics of network structure under cascading failure is challenging. \cite{Savla.Como.ea.TNSse14} considers the intermediate quasi-static regime, where, upon jump, the flow equilibrates on every link instantaneously, such that the flow ratio at every node is as determined by the distributed routing policy implemented at that node. \cite{Savla.Como.ea.TNSse14} provides an algorithm to synthesize routing policies which are maximally robust in this setting to perturbation \emph{processes} $\triangle(t)$, where $\triangle(t)$ is the cumulative reduction in capacity on the links until time $t$, and $\triangle(t)$ is non-decreasing in time.\footnote{The robustness results in Section~\ref{sec:capacity-perturbation-dynamic} extend to such a perturbation process, by replacing $\|\triangle\|_1$ there with $\|\triangle\|_{1,\infty}:=\lim_{t \to +\infty} \|\triangle(t)\|_1$.} 

The algorithm in \cite{Savla.Como.ea.TNSse14} implements a dynamic programming like computation over space, backwards from the sink node to the source node. At each stage, the algorithm computes the robust routing policy at a given node and the associated measure of robustness for the sub-network downstream of that node, and passes this information to its parent node. The routing policies synthesized by this algorithm, which performs only one iteration per node, can be shown to be maximally robust if the flow induced by it is monotonically non-decreasing on every active link during the cascading failure process. A simple sufficient condition for this is that the network is laterally symmetric about the source-sink pair, in terms of network structure as well as link capacities. Additionally, \cite{Savla.Como.ea.TNSse14} also provides a catalog of \emph{basis} networks, as well as rules to compose them into bigger networks, each of which ensures the monotone non-decreasing condition on flow. 


%% file: cascading-failure/cascading-failure-contagion.tex
\subsection{Contagion}
\label{sec:cascading-failure-contagion}
Consider the following alternate threshold-based cascade model from \cite{Blume.Easley.ea:11}, which is studied extensively in the context of social and biological contagion. Let each node $v$ choose a \emph{threshold} $\ell(v)$ independently from a distribution $\mu$ on $\naturals$. $\ell(v)$ represents the number of failed neighbors that $v$ can withstand before $v$ fails as well. The failure process works as follows. First declare all nodes with threshold 0 to have failed. Then, repeatedly check whether any node $v$ that has not yet failed has at least $\ell(v)$ failed neighbors -- if so, declare $v$ to have failed as well and continue iterating. $\mu$ can be thought of as determining the distribution of levels of ``health" of the nodes, and hence implicitly controlling the way the failure process spreads. 

For a given node $v$, let its \emph{failure probability} be denoted as $\failprob_{\mu}(v)$. Let $\failprob_{\mu}^* = \sup_{v \in \mc V} \failprob_{\mu} (v)$ be the \emph{maximum failure probability} in $\mc G$.  $\failprob_{\mu}^*$ can be viewed as a measure of resilience against cascading failures that operate under the threshold distribution $\mu$, and is also referred to as the $\mu$-risk of $\mc G$.
It is of interest to understand the relationship between $\mu$-risk and the structure of the underlying graph. 

\begin{figure}[htb!]
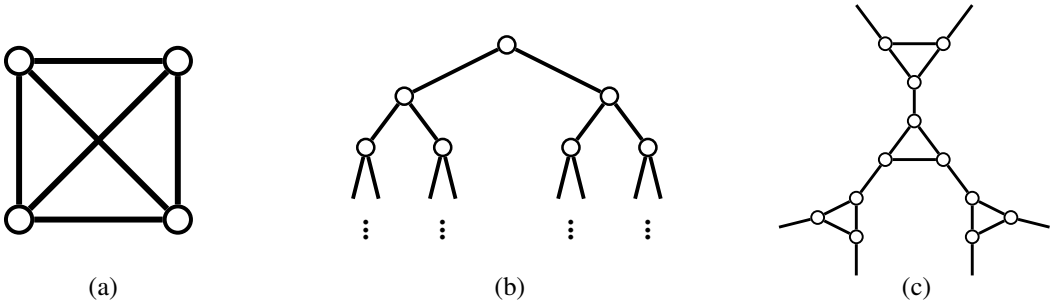

\begin{center}
\begin{minipage}[c]{.32\textwidth}
\begin{center}
\includestandalone[width=0.5\textwidth]{./fig/regular-tikz}
\end{center}
\end{minipage}
\begin{minipage}[c]{.32\textwidth}
\begin{center}
\includestandalone[width=0.8\textwidth]{./fig/binary-tree-tikz}
\end{center}
\end{minipage}
\begin{minipage}[c]{.32\textwidth}
\begin{center}
\includestandalone[width=0.7\textwidth]{./fig/tree-of-triangle-tikz}
\end{center}
\end{minipage}
\begin{minipage}[c]{.32\textwidth}
\begin{center}
(a)
\end{center}
\end{minipage}
\begin{minipage}[c]{.32\textwidth}
\begin{center}
(b)
\end{center}
\end{minipage}
\begin{minipage}[c]{.32\textwidth}
\begin{center}
(c)
\end{center}
\end{minipage}
\caption{(a) clique $K_{d+1}$ for $d=3$; (b) infinite complete $d$-ary tree $T_d$ for $d=2$ (c) tree of triangle $\triangletree_d$ for $d=3$.}
\label{fig:triangle-tree}
\end{center}
\end{figure}

While characterizing $\mu$-risk is challenging in general, certain graph classes lend themselves to useful insights. Within the class of $d$-regular graphs, denoted as $\graphclass_d$, contrasting the clique $K_{d+1}$ and the infinite complete $d$-ary tree $T_d$ (see Figure~\ref{fig:triangle-tree} for examples) shows that the $\mu$-risk minimizing graph structure depends on the distribution $\mu$.  However, at each $\mu$ with $0 < \mu(0) < 1$, at least one of $K_3$ or $T_2$ achieves strictly lower $\mu$-risk that every other graph in $\graphclass_2 \setminus \{K_3, T_2\}$. The behavior of $\mu$-risk on $\graphclass_d$ for $d>2$ is complicated: (i) there are distributions $\mu$ for which $K_{d+1}$ has strictly lower $\mu$-risk than any other $\mc G \in \graphclass_d$; (ii) for every $\mc G \in \graphclass_d$, there exists a $\mu_{\mc G}$ for which $T_d$ has a strictly lower $\mu_{\mc G}$-risk than $\mc G$; and (iii) there exist distribution $\mu$ for which the ($d$ regular) \emph{tree of triangles} $\triangletree_d$ (see Figure~\ref{fig:triangle-tree}(b) for an example), consisting essentially of a collection of disjoint triangles attached according to the structure of an infinite regular tree, has strictly lower $\mu$-risk than both $K_{d+1}$ and $T_d$. $\triangletree_d$ ``interpolates" between the complete neighborhood diversification of $T_d$ and the complete neighborhood closure of $K_{d+1}$.

The above results can be related to standard notions in relevant application domains. During epidemic disease, it is considered to be dangerous to belong to a large connected component, suggesting that $K_{d+1}$ is the most resilient network within $\graphclass_d$. On the other hand, a principle in financial networks is that it is important to have diversity among one's neighbors, i.e., lack of edges among one's neighbors, so that shocks are uncorrelated. This suggests $T_d$ to be the most resilient network in $\graphclass_d$. These observations can be formalized by appropriate $\mu$. Let $\left(\mu(0), \mu(1), \mu(2)\right) = \left(\varepsilon, x, 1 - \varepsilon - x \right)$ for some very small value of $\varepsilon>0$ and $\mu(j) =0$ for $j > 2$. If $x=1-\varepsilon$, i.e., thresholds are either $0$ or $1$, then a node's failure probability is strictly increasing in the size of the component it belongs to, and so $K_{d+1}$ uniquely minimizes the $\mu$-risk. On the other hand, there exists $x$ strictly between $0$ and $1-\varepsilon$, such that $x$ is very small, but significantly larger than $\varepsilon$, so that thresholds of $1$ are much more numerous than thresholds of $0$. In this case, $T_d$ is optimal, because, although failures are still rare, if a node $u$ has connected neighbors $v$ and $w$, then there is a non-trivial risk that $v$ will have threshold $0$ and $w$ will have threshold $1$, at which point $v$'s failure will ricochet off $w$ and bring down $u$ as well, even if $u$ has the maximum (and most likely) threshold of $2$. Therefore, in this scenario, it is safer to have no links among neighbors, even at the expense of producing very large connected components.  

%% file: fig/regular-tikz.tex
\begin{tikzpicture}

\path (-1,1) node(a) [circle,draw,line width=0.5mm] {$  $}
	(1,1) node(b) [circle,draw,line width=0.5mm] {$  $}
	(1,-1) node(c) [circle,draw,line width=0.5mm] {$  $}
	(-1,-1) node(d) [circle,draw,line width=0.5mm] {$  $};

\draw[very thick, line width=0.75mm] (a) -- (b) -- (c) -- (d) -- (a) -- (c);
\draw[very thick, line width=0.75mm] (d) -- (b);
\end{tikzpicture}

%% file: fig/binary-tree-tikz.tex
\begin{tikzpicture}

\path (0,0) node(a) [circle,draw,line width=0.5mm] {$  $}
	(-2,-1) node(b) [circle,draw,line width=0.5mm] {$  $}
	(2,-1) node(c) [circle,draw,line width=0.5mm] {$  $}
	(-2.75,-2) node(d) [circle,draw,line width=0.5mm] {$  $}
	(-1.25,-2) node(e) [circle,draw,line width=0.5mm] {$  $}
	(2.75,-2) node(f) [circle,draw,line width=0.5mm] {$  $}
	(1.25,-2) node(g) [circle,draw,line width=0.5mm] {$  $};

\draw[very thick, line width=0.75mm] (a) -- (b) -- (d) -- (-3,-3);
\draw[very thick, line width=0.75mm] (d) -- (-2.5,-3);
\draw[very thick, line width=0.75mm] (e) -- (-1.5,-3);
\draw[very thick, line width=0.75mm] (b) -- (e) -- (-1,-3);

\draw[very thick, line width=0.75mm] (a) -- (c) -- (f) -- (3,-3);
\draw[very thick, line width=0.75mm] (f) -- (2.5,-3);
\draw[very thick, line width=0.75mm] (g) -- (1.5,-3);
\draw[very thick, line width=0.75mm] (c) -- (g) -- (1,-3);

\draw (-2.75,-3.5) node {\Huge $\vdots$};
\draw (-1.25,-3.5) node {\Huge $\vdots$};
\draw (2.75,-3.5) node {\Huge $\vdots$};
\draw (1.25,-3.5) node {\Huge $\vdots$};

\end{tikzpicture}

%% file: fig/tree-of-triangle-tikz.tex
\begin{tikzpicture}

\path (-0.75,1) node(a) [circle,draw,line width=0.5mm] {$  $}
	(0.75,1) node(b) [circle,draw,line width=0.5mm] {$  $}
	(0,0) node(c) [circle,draw,line width=0.5mm] {$  $}
	(0,-1) node(d) [circle,draw,line width=0.5mm] {$  $}
	(-0.75,-2) node(e) [circle,draw,line width=0.5mm] {$  $}
	(0.75,-2) node(f) [circle,draw,line width=0.5mm] {$  $}
	(-1.5,-3) node(g) [circle,draw,line width=0.5mm] {$  $}
	(-2.5,-3.5) node(h) [circle,draw,line width=0.5mm] {$  $}
	(-1.5,-4) node(i) [circle,draw,line width=0.5mm] {$  $}
	(1.5,-3) node(j) [circle,draw,line width=0.5mm] {$  $}
	(2.5,-3.5) node(k) [circle,draw,line width=0.5mm] {$  $}
	(1.5,-4) node(l) [circle,draw,line width=0.5mm] {$  $};

\draw[very thick, line width=0.75mm] (-1.5,2) -- (a) -- (c) -- (d) -- (e) -- (g) -- (h) -- (-3.5,-3.75);
\draw[very thick, line width=0.75mm] (1.5,2) -- (b) -- (c);
\draw[very thick, line width=0.75mm] (d) -- (f) -- (j) -- (k) -- (3.5,-3.75);
\draw[very thick, line width=0.75mm] (g) -- (i) -- (-1.5,-5);
\draw[very thick, line width=0.75mm] (j) -- (l) -- (1.5,-5);
\draw[very thick, line width=0.75mm] (a) -- (b);
\draw[very thick, line width=0.75mm] (e) -- (f);
\draw[very thick, line width=0.75mm] (h) -- (i);
\draw[very thick, line width=0.75mm] (k) -- (l);
\end{tikzpicture}

%% file: asymptotic-static/asymptotic-static-main.tex
\section{Robustness of Asymptotically Large Networks: Static Setting}
\label{sec:asymptotic-static}
In Section~\ref{sec:cascading-failure-contagion}, we discussed the interplay between network structure and failure thresholds at nodes in determining robustness of financial networks under cascading failure. In this section, we discuss the interplay between network structure and interconnection weights in determining robustness under an equilibrium model in economic networks. 
Specifically, it is of interest to understand how idiosyncratic shocks at nodes translate into fluctuations of a meaningful aggregate quantity associated with a network. A central limit theorem type argument suggests that if the shocks are independent, then fluctuations in the quantity corresponding to simple summation of the shocks would decay proportional to $1/\sqrt{n}$. However, the interconnections induced by the network can function as a potential propagation mechanism, under which the $1/\sqrt{n}$ decay may not hold true. The analysis of such propagation mechanisms and their dependence on network structure, in the context of inter-sectoral economic networks is provided in \cite{Acemoglu.Carvalho.ea:12,Acemoglu.Ozdaglar.ea:17}.

Let the nodes in $\mc V$ represent sectors that produce goods consumed by a representative household and the household provides a fixed inelastic cumulative unit labor for all the sectors. Let $\ell_i$ be the (fraction of) labor consumed by node $i$, and let $x_i$ be the amount of good produced by node $i$. The household utility function is of Cobb-Douglas type, i.e., it is proportional to $\Pi_{i \in [n]} x_i^{\eta_i}$, where 
$\eta_i>0$ is $i$'s share in the household's utility function, normalized such that $\sum_{i \in [n]} \eta_i=1$. The amount of good produced by node $i$ is modeled as:
%
%
\begin{equation}
\label{eq:node-production}
x_i = \shock_i^{\alpha} \ell_i^{\alpha} \Pi_{j=1}^n x_{ij}^{(1-\alpha) w_{ij}}
\end{equation}
where $x_{ij}$ is the amount of good $j$ used in the production of good $i$, $\alpha \in (0,1)$ and $w_{ij} \in (0,1)$ are the shares of labor and good $j$, respectively, in the input for production by $i$, satisfying $\sum_j w_{ij} = 1$ for all $i$, and $\shock_i$ is the idiosyncratic productivity shock to sector $i$.   
At the competitive equilibrium, i.e., where the household maximizes utility, the individual sectors maximize profits, and labor and commodity market clears, the logarithm of the aggregate output of the network is given by 
\begin{equation}
\label{eq:aggregate-output}
y=v^T \omega
\end{equation}
 where $\omega = [\omega_1, \ldots, \omega_n]^T$, with $\omega_i=\log(\shock_i)$, is the vector of \emph{microscopic shocks}, and 
the $i$-th component of the \emph{influence vector} $v=[v^{(1)}, \ldots, v^{(n)}]^T$, also referred to as the \emph{Domar weight} of node $i$ in economic network context, is given by
\begin{equation}
\label{eq:influence-vector-def-general}
v^{(i)} = \sum_{j \in [n]} \eta_j L_{ji}
\end{equation}
where $L_{ji}$ is the $(j,i)$-element of the Leontief inverse $L=\left(I - (1-\alpha) \mc W \right)^{-1}$.
That is, the logarithm of aggregate output is a linear combination of log node shocks with coefficients determined by the elements of the influence vector. 
%
The node shocks, and hence $\omega_i$, are modeled to be random variables independent across the nodes. If $\mathbb{E}(\omega_i)=0$ for all $i \in [n]$, then $\mathbb{E}(y)=0$. It is of interest to study the standard deviation, also referred to as \emph{aggregate volatility}, of $y$. This is done in Section~\ref{subsec:aggregate-fluctuations}. It is also of interest to study \emph{aggregate output's $\tau$-tail ratio} (cf. \eqref{eq:tail-ratio}).
%
A network is said to exhibit macroscopic tail risk if the aggregated output associated with it satisfies $\lim_{\tau \to \infty} r_y(\tau)=0$. This is studied in Section~\ref{sec:tail-risk}.

It can be shown that, for any finite network with $n$ nodes, if $\{\omega_i\}_{i \in [n]}$ exhibit tail risks, then the network exhibits macroscopic tail risk as well. Therefore, for meaningful analysis, one considers a sequence of networks $\{\mc G_n=(\mc V_n, \mc E_n, \mc W_n)\}_{n \in \naturals}$ along with a collection of distributions of log node shocks $\{F_{in}\}_{i \in \mc V_n, n \in \naturals}$, and studies aggregate volatility and tail risk as $n \to \infty$. We let the corresponding output vectors, influence vectors, and tail ratios be denoted as $\{y_n\}_{n \in \naturals}$, $\{v_n\}_{n \in \naturals}$ and $\{r_n(\tau_n)\}_{n \in \naturals}$ respectively.


\input{./asymptotic-static/aggregate-fluctuations}

\input{./asymptotic-static/tail-risk}

%% file: asymptotic-static/aggregate-fluctuations.tex
\subsection{Aggregate Volatility}
\label{subsec:aggregate-fluctuations}
Let the variances $\var[\omega_{in}]$ be finite and uniformly strictly bounded away from zero to be able to focus on the effects of network structure as $n \to \infty$. Consider the special case when $\eta_i=1/n$ for all $i \in [n]$. In this case, \eqref{eq:influence-vector-def-general} specializes to:
\begin{equation}
\label{eq:influence-vector-def}
v_n=\frac{1}{n} \left[I - (1-\alpha) \mc W^T \right]^{-1} \onebf
\end{equation}
For a given $n$, \eqref{eq:aggregate-output} implies 
\begin{equation}
\label{eq:output-normalized}
(\var[y_n])^{1/2}=\Theta(\|v_n\|_2)
\end{equation} 

\begin{figure}[htb!]
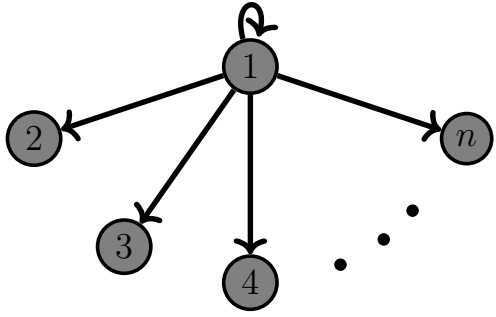

\begin{center}
\includestandalone[width=0.4\textwidth]{./fig/star-network-tikz}
\end{center}
\caption{A network where one sector (node 1) is the only supplier.}
\label{fig:star}
\end{figure}
It is of interest to study the dependence of volatility on $n$, and   
compare against the $1/\sqrt{n}$ behavior as would be implied by the naive application of the central limit theorem. One extreme case is the network depicted in Figure~\ref{fig:star}, for which $\|v_n\|_2=\Theta(1)$, i.e., volatility does not vanish even as $n \to \infty$. However, it is natural to expect that in most realistic situations, volatility vanishes as $n \to \infty$. Indeed, one can show the following results that are complementary to \eqref{eq:output-normalized}, when all the $\{\omega_{in}\}_{i,n}$ have the same variance $\sigma^2$. If all $\{\omega_{in}\}_{i,n}$ are normally distributed, then $\frac{1}{\|v_n\|_2} y_n \xrightarrow{d} \mc N(0,\sigma^2)$. The same convergence result also holds true in general if there exists a cumulative distribution function $\bar{F}$ such that $F_{in}(x) < \bar{F}(x)$ for $x < -a$ and $F_{in}(x) > \bar{F}(x)$ for $x>a$ for some $a>0$, and if $\frac{\|v_n\|_{\infty}}{\|v_n\|_2} \to 0$. The last condition captures the dependence of network structure and weights through the influence vector $v_n$, and implies that $\|v_n\|_{\infty}$, which captures the influence of the most central node, converges to zero faster than $\|v_n\|_{2}$.


A lower bound on volatility can be obtained in terms of the outdegrees $d_1^n, d_2^n, \ldots, d_n^n$, where recall from Section~\ref{sec:notations} that $d_i^n:=\sum_{j=1}^n w_{ij}$ for all $i \in [n]$:
\begin{equation}
\label{eq:volatility-lower-bound-1}
\left(\var[y_n] \right)^{1/2} = \Omega \left(\frac{1+\text{CV}_n}{\sqrt{n}} \right)
\end{equation}
where the coefficient of variation $\text{CV}_n$ measures the extent of asymmetry between nodes: 
$$
\text{CV}_n := \frac{1}{\bar{d}_n} \left(\frac{1}{n-1} \sum_{i=1}^n \left(d^n_i - \bar{d}_n \right)^2 \right)^{1/2}
$$
where $\bar{d}_n=\sum_{i=1}^n d_i^n/n$ is the average outdegree. \eqref{eq:volatility-lower-bound-1} implies that asymmetry can cause the volatility to decay slower than $1/\sqrt{n}$. For example, for the network depicted in Figure~\ref{fig:star}, $
\text{CV}_n=\Theta(\sqrt{n})$, which then implies that the aggregate volatility is lower bounded by a constant for all values of $n$. More generally, if the network contains a ``dominant" node whose degree grows linearly with $n$, then the aggregate volatility remains bounded away from zero. A complementary result holds true for networks whose degree sequences have ``heavier tails", or formally, for a sequence of networks having a \emph{power law degree sequence}, i.e., if there exist a constant $\beta>1$, a slowly varying function $L$ satisfying $\lim_{t \to \infty} L(t) t^{\delta}=\infty$ and $\lim_{t \to \infty} L(t) t^{-\delta}=0$ for all $\delta>0$, and a sequence of positive numbers $\gamma_n = \Theta(1)$, such that, for all $n \in \naturals$ and all $k < d_{\max}^n = \Theta(n^{1/\beta})$, with $d_{\max}^n$ being the maximum outdegree of $\mc G_n$, we have the empirical counter cumulative distribution function $P_n(k) \equiv \frac{1}{n}| \{i \in \mc V_n: \, d_i^n >k\}|$ satisfying $P_n(k)=\gamma_n k^{-\beta} L(k)$. For such networks with $\beta \in (1,2)$, the aggregate volatility can be shown to be $\Omega(n^{-(\beta-1)/\beta - \delta})$ for arbitrary $\delta>0$. 

\begin{figure}[htb!]
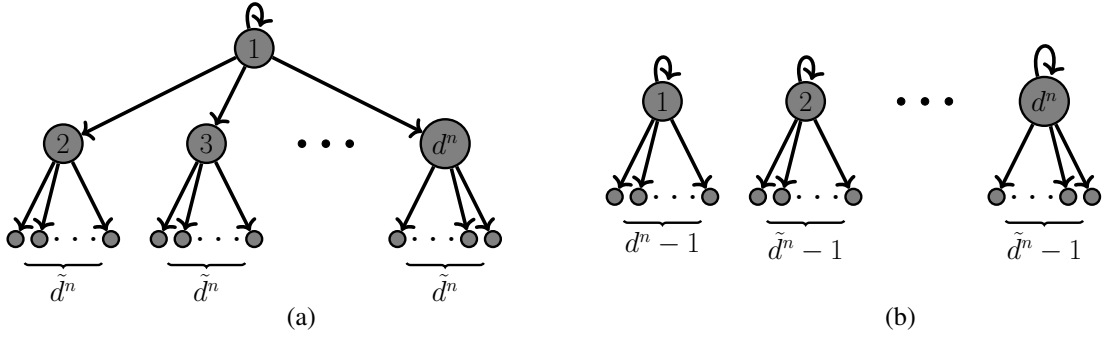

\begin{center}
\begin{minipage}[c]{.475\textwidth}
\includestandalone[width=0.85\textwidth]{./fig/second-order-motivation-a-tikz}
\end{minipage}
\begin{minipage}[c]{.475\textwidth}
\includestandalone[width=0.85\linewidth]{./fig/second-order-motivation-b-tikz} 
\end{minipage}
\begin{minipage}[c]{.475\textwidth}
\begin{center}
(a)
\end{center}
\end{minipage}
\begin{minipage}[c]{.475\textwidth}
\begin{center}
(b)
\end{center}
\end{minipage}
\end{center}
\caption{Networks with the same degree sequence but with different aggregate volatilities~\cite{Acemoglu.Carvalho.ea:12}.}
\label{fig:second-order-motivation}
\end{figure}

The above results on dependence of volatility on the degree properties of a network capture \emph{first order} effects of network structure and may not be sufficient to identify network properties under which the volatility decays slower than $1/\sqrt{n}$. For example, 
both the networks shown in Figure~\ref{fig:second-order-motivation} have the same degree sequence: node labeled $1$ with degree $d^n=\Theta(\sqrt{n})$, nodes $2, \ldots, d^n$ with some degree $\tilde{d}^n$, and other nodes with degree zero. However, the aggregate volatilities for the networks in (a) and (b), respectively, scale as $\Theta(1)$ and $\Theta(1/\sqrt[4]{n})$, independent of $\tilde{d}^n$. The first-order interconnections provide little or no information on the extent of \emph{cascade} effects, whereby shocks to a node affect only its immediate outgoing nodes, but also outgoing nodes of those nodes, and so on. The \emph{second-order interconnectivity coefficient} captures such effects:  
$$
\tau_2(\mc W_n) := \sum_{i=1}^n \sum_{j \neq i} \sum_{k \neq i,j} w^n_{ji} \, w^n_{ki} \, d^n_j \, d^n_k
$$
$\tau_2$ takes higher values when high-degree nodes share incoming nodes with other high-degree nodes, as opposed to low-degree ones. The bound in \eqref{eq:volatility-lower-bound-1} can then be strengthened as:
\begin{equation}
\label{eq:volatility-lower-bound-2}
\left(\var[y_n] \right)^{1/2} = \Omega \left(\frac{1+\text{CV}_n}{\sqrt{n}} + \frac{\sqrt{\tau_2(\mc W_n)}}{n}\right)
\end{equation}

The effect of second order interconnections can also be formalized in terms of \emph{second-order degree} sequence, where the second-order degree of a node $i$ is defined as the weighted sum of the degrees of the nodes which are outgoing from $i$, i.e., $\sum_{j \in [n]} d_j^n w_{ji}^n$. If the second-order degree sequence associated with $\{\mc G_n\}_{n \in \naturals}$ has power law tail with shape parameter $\zeta \in (1,2)$, then the aggregate volatility satisfies $(\var[y_n])^{1/2} = \Omega(n^{-(\zeta-1)/\zeta - \delta})$ for any $\delta>0$.

\begin{figure}[htb!]
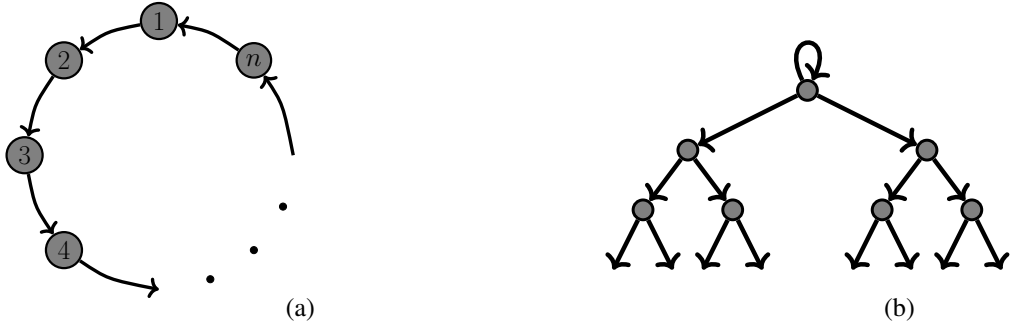

\begin{center}
\begin{minipage}[c]{.475\textwidth}
\includestandalone[width=0.5\linewidth]{./fig/ring-tikz} 
\end{minipage}
\begin{minipage}[c]{.475\textwidth}
\includestandalone[width=0.7\linewidth]{./fig/binary-tree-directed-tikz} 
\end{minipage}
\begin{minipage}[c]{.475\textwidth}
\begin{center}
(a)
\end{center}
\end{minipage}
\begin{minipage}[c]{.475\textwidth}
\begin{center}
(b)
\end{center}
\end{minipage}
\end{center}
\caption{Examples of balanced networks: (a) ring (b) binary tree.}
\label{fig:balanced-network}
\end{figure}

\eqref{eq:volatility-lower-bound-2} can be strengthened further to capture asymmetry in higher-order interconnections. Conversely, if the node degrees have limited variation, then the volatility decays at $1/\sqrt{n}$. Formally, for a  sequence of balanced networks $\{\mc G_n\}_{n \in \naturals}$, i.e., ones satisfying $\max_{i \in [n]} d_i^n = \Theta(1)$, there exists $\bar{\alpha} \in (0,1)$, such that for $\alpha \geq \bar{\alpha}$, $(\var[y_n])^{1/2}=\Theta(1/\sqrt{n})$, where we recall from \eqref{eq:node-production} that $\alpha$ denotes the share of labor in the production of goods by nodes. Simple examples of balanced networks are illustrated in Figure~\ref{fig:balanced-network}.

%% file: fig/star-network-tikz.tex
\begin{tikzpicture}

\path (0,0) node(a) [circle,fill=gray,draw,line width=0.5mm] {\Large $1$}
(-3,-1) node(b) [circle,fill=gray,draw,line width=0.5mm] {\Large $2$}
(-1.75,-2.5) node(c) [circle,fill=gray,draw,line width=0.5mm] {\Large $3$}
(0,-3) node(d) [circle,fill=gray,draw,line width=0.5mm] {\Large $4$}
(3,-1) node(e) [circle,fill=gray,draw,line width=0.5mm] {\Large $n$};

\path (a) edge [loop above, line width=0.75mm] (0,0.75) (a);

\draw[very thick, ->, line width=0.75mm] (a) -- (b);
\draw[very thick, ->, line width=0.75mm] (a) -- (c);
\draw[very thick, ->, line width=0.75mm] (a) -- (d);
\draw[very thick, ->, line width=0.75mm] (a) -- (e);

\draw[fill] (1.25,-2.75) circle (.5ex);
\draw[fill] (1.85,-2.4) circle (.5ex);
\draw[fill] (2.25,-2) circle (.5ex);

\end{tikzpicture}

%% file: fig/second-order-motivation-a-tikz.tex
\begin{tikzpicture}

\path (0,0) node(a) [circle,fill=gray,draw,line width=0.5mm] {\LARGE $1$}
(-4,-2) node(b) [circle,fill=gray,draw,line width=0.5mm] {\LARGE $2$}
(-1,-2) node(c) [circle,fill=gray,draw,line width=0.5mm] {\LARGE $3$}
(4,-2) node(d) [circle,fill=gray,draw,line width=0.5mm] {\LARGE $d^n$}
(-5,-4) node(e) [circle,fill=gray,draw,line width=0.5mm] {\LARGE $ $}
(-4.5,-4) node(f) [circle,fill=gray,draw,line width=0.5mm] {\LARGE $ $}
(-3,-4) node(g) [circle,fill=gray,draw,line width=0.5mm] {\LARGE $ $}
(-2,-4) node(h) [circle,fill=gray,draw,line width=0.5mm] {\LARGE $ $}
(-1.5,-4) node(i) [circle,fill=gray,draw,line width=0.5mm] {\LARGE $ $}
(0,-4) node(j) [circle,fill=gray,draw,line width=0.5mm] {\LARGE $ $}
(3,-4) node(k) [circle,fill=gray,draw,line width=0.5mm] {\LARGE $ $}
(4.5,-4) node(l) [circle,fill=gray,draw,line width=0.5mm] {\LARGE $ $}
(5,-4) node(m) [circle,fill=gray,draw,line width=0.5mm] {\LARGE $ $}
;

\path (a) edge [loop above, line width=0.75mm] (0,0.75) (a);

%
\draw[very thick, ->, line width=0.75mm] (a) -- (b);
\draw[very thick, ->, line width=0.75mm] (a) -- (c);
\draw[very thick, ->, line width=0.75mm] (a) -- (d);

\draw[very thick, ->, line width=0.75mm] (b) -- (e);
\draw[very thick, ->, line width=0.75mm] (b) -- (f);
\draw[very thick, ->, line width=0.75mm] (b) -- (g);

\draw[very thick, ->, line width=0.75mm] (c) -- (h);
\draw[very thick, ->, line width=0.75mm] (c) -- (i);
\draw[very thick, ->, line width=0.75mm] (c) -- (j);

\draw[very thick, ->, line width=0.75mm] (d) -- (k);
\draw[very thick, ->, line width=0.75mm] (d) -- (l);
\draw[very thick, ->, line width=0.75mm] (d) -- (m);

\draw[fill] (1.5,-2) circle (.5ex);
\draw[fill] (1,-2) circle (.5ex);
\draw[fill] (2,-2) circle (.5ex);

\draw (-3.75,-4) node {\Huge $\ldots$};
\draw (-0.75,-4) node {\Huge $\ldots$};
\draw (3.75,-4) node {\Huge $\ldots$};

\draw [
    very thick, line width = 0.5mm,
    decoration={
        brace,
        mirror,
        raise=0.5cm
    },
    decorate
] (e) -- (g) 
node [pos=0.5,anchor=north,yshift=-0.6cm] {\LARGE $\tilde{d}^n$}; 

\draw [
    very thick, line width = 0.5mm,
    decoration={
        brace,
        mirror,
        raise=0.5cm
    },
    decorate
] (h) -- (j) 
node [pos=0.5,anchor=north,yshift=-0.6cm] {\LARGE $\tilde{d}^n$}; 

\draw [
    very thick, line width = 0.5mm,
    decoration={
        brace,
        mirror,
        raise=0.5cm
    },
    decorate
] (k) -- (m) 
node [pos=0.5,anchor=north,yshift=-0.6cm] {\LARGE $\tilde{d}^n$}; 

\end{tikzpicture}

%% file: fig/second-order-motivation-b-tikz.tex
\begin{tikzpicture}

\path 
(-4,-2) node(b) [circle,fill=gray,draw,line width=0.5mm] {\LARGE $1$}
(-1,-2) node(c) [circle,fill=gray,draw,line width=0.5mm] {\LARGE $2$}
(4,-2) node(d) [circle,fill=gray,draw,line width=0.5mm] {\LARGE $d^n$}
(-5,-4) node(e) [circle,fill=gray,draw,line width=0.5mm] {\LARGE $ $}
(-4.5,-4) node(f) [circle,fill=gray,draw,line width=0.5mm] {\LARGE $ $}
(-3,-4) node(g) [circle,fill=gray,draw,line width=0.5mm] {\LARGE $ $}
(-2,-4) node(h) [circle,fill=gray,draw,line width=0.5mm] {\LARGE $ $}
(-1.5,-4) node(i) [circle,fill=gray,draw,line width=0.5mm] {\LARGE $ $}
(0,-4) node(j) [circle,fill=gray,draw,line width=0.5mm] {\LARGE $ $}
(3,-4) node(k) [circle,fill=gray,draw,line width=0.5mm] {\LARGE $ $}
(4.5,-4) node(l) [circle,fill=gray,draw,line width=0.5mm] {\LARGE $ $}
(5,-4) node(m) [circle,fill=gray,draw,line width=0.5mm] {\LARGE $ $}
;

\path (b) edge [loop above, line width=0.75mm] (-4,-1.25) (b);
\path (c) edge [loop above, line width=0.75mm] (-1,-1.25) (c);
\path (d) edge [loop above, line width=0.75mm] (4,-1.25) (d);

\draw[very thick, ->, line width=0.75mm] (b) -- (e);
\draw[very thick, ->, line width=0.75mm] (b) -- (f);
\draw[very thick, ->, line width=0.75mm] (b) -- (g);

\draw[very thick, ->, line width=0.75mm] (c) -- (h);
\draw[very thick, ->, line width=0.75mm] (c) -- (i);
\draw[very thick, ->, line width=0.75mm] (c) -- (j);

\draw[very thick, ->, line width=0.75mm] (d) -- (k);
\draw[very thick, ->, line width=0.75mm] (d) -- (l);
\draw[very thick, ->, line width=0.75mm] (d) -- (m);

\draw[fill] (1.5,-2) circle (.5ex);
\draw[fill] (1,-2) circle (.5ex);
\draw[fill] (2,-2) circle (.5ex);

\draw (-3.75,-4) node {\Huge $\ldots$};
\draw (-0.75,-4) node {\Huge $\ldots$};
\draw (3.75,-4) node {\Huge $\ldots$};

\draw [
    very thick, line width = 0.5mm,
    decoration={
        brace,
        mirror,
        raise=0.5cm
    },
    decorate
] (e) -- (g) 
node [pos=0.5,anchor=north,yshift=-0.6cm] {\LARGE $d^n-1$}; 

\draw [
    very thick, line width = 0.5mm,
    decoration={
        brace,
        mirror,
        raise=0.5cm
    },
    decorate
] (h) -- (j) 
node [pos=0.5,anchor=north,yshift=-0.6cm] {\LARGE $\tilde{d}^n-1$}; 

\draw [
    very thick, line width = 0.5mm,
    decoration={
        brace,
        mirror,
        raise=0.5cm
    },
    decorate
] (k) -- (m) 
node [pos=0.5,anchor=north,yshift=-0.6cm] {\LARGE $\tilde{d}^n-1$}; 

\end{tikzpicture}

%% file: fig/ring-tikz.tex
\begin{tikzpicture}

\path 
(0,3) node(a) [circle,fill=gray,draw,line width=0.5mm] {\LARGE $1$}
	(-3*1.414/2,3*1.414/2) node(b) [circle,fill=gray,draw,line width=0.5mm] {\LARGE $2$}
	(-3,0) node(c) [circle,,fill=gray,draw,line width=0.5mm] {\LARGE $3$}
	(-3*1.414/2,-3*1.414/2) node(d) [circle,fill=gray,draw,line width=0.5mm] {\LARGE $4$}
	(3*1.414/2,3*1.414/2) node(e) [circle,fill=gray,draw,line width=0.5mm] {\LARGE $n$}
	;

\draw[very thick, ->, line width=0.75mm] (a) .. controls (-3*0.3827,3*0.9239) .. (b);
\draw[very thick, ->, line width=0.75mm] (b) .. controls (-3*0.9239,3*0.3827) .. (c);
\draw[very thick, ->, line width=0.75mm] (c) .. controls (-3*0.9239,-3*0.3827) .. (d);
\draw[very thick, ->, line width=0.75mm] (d) .. controls (-3*0.3827,-3*0.9239) .. (0,-3);
\draw[very thick, ->, line width=0.75mm] (3,0) .. controls (3*0.9239,3*0.3827) .. (e);
\draw[very thick, ->, line width=0.75mm] (e) .. controls (3*0.3827,3*0.9239) .. (a);

\draw[fill] (3*0.3827,-3*0.9239) circle (.5ex);
\draw[fill] (3*1.414/2,-3*1.414/2) circle (.5ex);
\draw[fill] (3*0.9239,-3*0.3827) circle (.5ex);

\end{tikzpicture}

%% file: fig/binary-tree-directed-tikz.tex
\begin{tikzpicture}

\path (0,0) node(a) [circle,fill=gray,draw,line width=0.5mm] {\Huge $  $}
	(-2,-1) node(b) [circle,fill=gray,draw,line width=0.5mm] {$  $}
	(2,-1) node(c) [circle,fill=gray,draw,line width=0.5mm] {$  $}
	(-2.75,-2) node(d) [circle,fill=gray,draw,line width=0.5mm] {$  $}
	(-1.25,-2) node(e) [circle,fill=gray,draw,line width=0.5mm] {$  $}
	(2.75,-2) node(f) [circle,fill=gray,draw,line width=0.5mm] {$  $}
	(1.25,-2) node(g) [circle,fill=gray,draw,line width=0.5mm] {$  $};

\draw[very thick, ->, line width=0.75mm] (a) -- (b); 
\draw[very thick, ->, line width=0.75mm] (b) -- (d);
\draw[very thick, ->, line width=0.75mm] (d) -- (-3.25,-3);
\draw[very thick, ->, line width=0.75mm] (d) -- (-2.25,-3);
\draw[very thick, ->, line width=0.75mm] (e) -- (-1.75,-3);
\draw[very thick, ->, line width=0.75mm] (b) -- (e);
\draw[very thick, ->, line width=0.75mm] (e) -- (-0.75,-3);

\draw[very thick, ->, line width=0.75mm] (a) -- (c);
\draw[very thick, ->, line width=0.75mm] (c) -- (f);
\draw[very thick, ->, line width=0.75mm] (f) -- (3.25,-3);
\draw[very thick, ->, line width=0.75mm] (f) -- (2.25,-3);
\draw[very thick, ->, line width=0.75mm] (g) -- (1.75,-3);
\draw[very thick, ->, line width=0.75mm] (c) -- (g);
\draw[very thick, ->, line width=0.75mm] (g) -- (0.75,-3);

\draw[very thick, ->, line width=0.75mm] (a) .. controls (-0.5,1) and (0.5,1) .. (a);

\end{tikzpicture}

%% file: asymptotic-static/tail-risk.tex
\subsection{Macroscopic Tail Risks}
\label{sec:tail-risk} 
\ksmarginsoft{Discuss specific graph structures to illustrate results in Sections~\ref{subsec:aggregate-fluctuations} and \ref{sec:tail-risk}}

Since the notion of macroscopic tail risk involves $\tau \to \infty$ argument, the order in which $\tau$ and $n$ are taken to infinity becomes crucial. We index $\tau$ in terms of $n$ as $\{\tau_n\}_{n \in \naturals}$ to highlight the dependence of the rates at which the two limits are taken. The inspiration for the correct dependence of $\tau$ on $n$ comes from the following result for \emph{simple} economic networks, i.e., when $\alpha=1$ and $\eta_i=1/n$ for all $i \in [n]$ (cf. the formulation for \eqref{eq:node-production}). If $\lim_{n \to \infty} \tau_n/\sqrt{n}=0$, then $\lim_{n \to \infty} r_n(\tau_n)=1$ for all light-tailed microscopic shocks; if $\lim_{n \to \infty} \tau_n/\sqrt{n}=\infty$, then there exist light-tailed microscopic shocks such that $\lim_{n \to \infty} r_n(\tau_n)=0$. The latter in particular contradicts a standard argument under which $\{\omega_i\}_{i \in [n]}$ should have no aggregate impact as $n \to \infty$ for simple economic networks if the rate of growth of $\tau_n$ is fast enough, as noted in \cite{Acemoglu.Ozdaglar.ea:17}. Accordingly, we say that a sequence of, not necessarily simple, networks exhibits macroscopic tail risks if $\lim_{n \to \infty} r_n(c \sqrt{n})=0$ for all $c>0$.

The presence or absence of macroscopic tail risks is determined by the interaction between the extent of heterogeneity in the Domar weights in \eqref{eq:influence-vector-def-general} and the distribution of microscopic shocks $\{\omega_{in}\}$. For instance, if the microscopic shocks are normally distributed, then no sequence of networks exhibits macroscopic tail risks. In more interesting cases, a measure of node dominance of a given network:
\begin{equation}
\label{eq:delta-vector}
\delta^n_v:=\frac{\|v_n\|_{\infty}}{\|v_n\|_2 /\sqrt{n}}
\end{equation}
plays a key role in macroscopic tail risks. The normalization factor $\sqrt{n}$ is meant to reflect that $\delta^n_v$ captures dominance relative to simple networks, where $\|v_n\|_{\infty}=1/n$ and $\|v_n\|_2=1/\sqrt{n}$, and therefore the node dominance is 1 for simple networks. 

\begin{figure}[htb!]
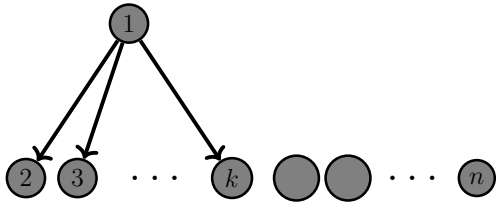

\begin{center}
\includestandalone[width=0.4\textwidth]{./fig/tail-risk-diminishing-connectivity-tikz}
\end{center}
\caption{A network where one sector (node 1) is the only supplier, and only a subset of $k-1$ nodes require input.}
\label{fig:diminishing}
\end{figure}

Under \emph{exponential-tailed} shocks, a sequence of networks exhibits macroscopic tail risks if and only if $\lim_{n \to \infty} \delta^n_v = \infty$. A shock has an exponential tail if its cumulative distribution function satisfies $\lim_{x \to \infty} \frac{1}{x} \log F(-x) = - \gamma$ for some $\gamma>0$. For example, this is satisfied if, for some polynomial function $Q(x)$, $F(-x)=1-F(x)=Q(x) e^{-\gamma x}$ for all $x \geq 0$. While exponential-tailed shocks are light-tailed distributions, they exhibit tail risks. Therefore, heterogeneity in entries of the influence vector is essential not only in generating aggregate volatility, as in Section~\ref{subsec:aggregate-fluctuations}, but also in translating microscopic tail risk into macroscopic tail risks. However, the role played in these two aspects are fundamentally distinct. For example, under exponential tailed microscopic shocks, a sequence of networks for which $\lim_{n \to \infty} \delta_v^n/\sqrt{n}=0$ and $\lim_{n \to \infty} \delta_v^n = \infty$ exhibits macroscopic tail risks, even though aggregate output is asymptotically normally distributed. As an illustration, consider the network in Figure~\ref{fig:diminishing} with $\eta_i=1/n$ for all $i \in [n]$. One can verify that, in this case, $\lim_{n \to \infty} \delta_v^n = \infty$, and hence has macroscopic tail risk under exponentially-tailed shocks if and only if $k \to \infty$ as $n \to \infty$. The latter could happen even if the only supply node is connected to a diminishing fraction of other nodes, e.g., $k = \log n$ which satisfies $\lim_{n \to \infty} k/n=0$, which in turn can be shown to imply $\frac{\|v_n\|_{\infty}}{\|v_n\|_2} \to 0$, and hence no aggregate volatility from the discussion in Section~\ref{subsec:aggregate-fluctuations}.

The above results can be generalized to a larger subclass of light-tailed microscopic shocks, such as the ones with \emph{super-exponential} distributions with shape parameter $\nu \in (1,2)$, in the sense that $\lim_{x \to \infty} \frac{1}{x^{\nu}} \log F(-x) = -\gamma$ for some $\gamma>0$. 
For example, this is satisfied if $F(-x)=1-F(x)=Q(x) \exp(-\gamma x^{\nu})$ for some polynomial function $Q(x)$. Such distributions exhibit tail risks while having tails that are lighter than that of the exponential distribution. Under microscopic shocks with super-exponential tails with shape parameter $\nu \in (1,2)$, a sequence of networks exhibits no macroscopic tail risk if $\liminf_{n \to \infty} \delta_v^n < \infty$, whereas it exhibits macroscopic tail risk if $\lim_{n \to \infty} \delta_v^n/n^{(\nu-1)/\nu}=\infty$.

Expectedly, one gets macroscopic tail risks for heavy-tailed shocks. Specifically, if the microeconomic shocks are Pareto (heavy-)tailed, i.e., if $\lim_{x \to \infty} \frac{1}{\log x} \log F(-x)=-\lambda$, with $\lambda>2$, then any sequence of networks exhibits macroeconomic tail risks. This is because the likelihood that at least one node is hit with a large shock is high. However, as we discussed already in this section that macroeconomic tail risks can emerge not just due to micro shocks that are drawn from heavy-tailed distributions, but also as a consequence of the interplay between relatively light-tailed distributions and heterogeneity in Domar weights. In fact, for a sequence of simple networks subject to Pareto tailed shocks, there exist a sequence of networks subject to exponential-tailed shocks which exhibits an identical level of macroeconomic tail risks~\cite{Acemoglu.Ozdaglar.ea:17}.  

\paragraph{Future Research Directions}
The results reviewed in this section depend on the linear relationship between Domar weights $v$ and external shocks $\omega$ in \eqref{eq:aggregate-output}, and on the specific dependence of $v$ on $\mc W$ in \eqref{eq:influence-vector-def-general}. This in turn depends on the underlying specific equilibrium setup. \cite{Acemoglu.Ozdaglar.ea:17} suggests practical settings under which $\mc W$ is not independent of $\omega$, which then induces nonlinear relationship between $v$ and $\omega$. Investigating aggregate volatility and macroscopic tail risks under such nonlinearities would be interesting.

%% file: fig/tail-risk-diminishing-connectivity-tikz.tex
\begin{tikzpicture}

\path 
(0,0) node(a) [circle,fill=gray,draw,line width=0.5mm] {\Large $1$}
(-2,-3) node(b) [circle,fill=gray,draw,line width=0.5mm] {\Large $2$}
(-1,-3) node(c) [circle,fill=gray,draw,line width=0.5mm] {\Large $3$}
(2,-3) node(d) [circle,fill=gray,draw,line width=0.5mm] {\Large $k$}
(6.75,-3) node(g) [circle,fill=gray,draw,line width=0.5mm] {\Large $n$}
;

\draw[fill=gray,draw,line width=0.5mm] (3.25,-3) circle (2.75ex);
\draw[fill=gray,draw,line width=0.5mm] (4.25,-3) circle (2.75ex);

\draw (0.5,-3) node {\Huge $\ldots$};
\draw (5.5,-3) node {\Huge $\ldots$};

\draw[very thick, ->, line width=0.75mm] (a) -- (b);
\draw[very thick, ->, line width=0.75mm] (a) -- (c);
\draw[very thick, ->, line width=0.75mm] (a) -- (d);

\end{tikzpicture}

%% file: asymptotic-dynamic/consensus.tex
\section{Robustness of Asymptotically Large Networks: Dynamical Setting}
\label{sec:asymptotic-dynamical}
The notion of aggregate output from Section~\ref{sec:asymptotic-static} can be related to the following dynamics:
\begin{equation}
\label{eq:dynamics-asymptotic}
x(t+1) = A x(t) + \omega \delta(0,t), \quad t \in \{0, 1, 2, \ldots\}, \qquad x(0)=0
\end{equation}
where $x(t)$ is the vector of state variables, $A$ is the $n \times n$ state transition matrix which is assumed to be Schur stable, i.e., it satisfies $\rho(A)<1$, 
$\delta(0,t)$ is the Kronecker delta function, and $\omega \in \reals^n$ is exogenous to the system. Under this dynamics, since $x(t)=A^{t-1} \omega$, we have $x_{\infty}:=\frac{1}{n}\sum_{t=0}^{\infty} \onebf^T x(t)=\frac{1}{n}\onebf^T\left(I + A + A^2 + \ldots \right) \omega=\frac{1}{n}\onebf^T(I-A)^{-1} \omega$. Comparing with \eqref{eq:aggregate-output} and \eqref{eq:influence-vector-def}, $x_{\infty}$ is the aggregate output when each node's share in household's utility is identical, i.e., $\eta_i = 1/n$ for all $i \in [n]$, and $A=(1-\alpha) \mc W$. 

\begin{figure}[htb!]
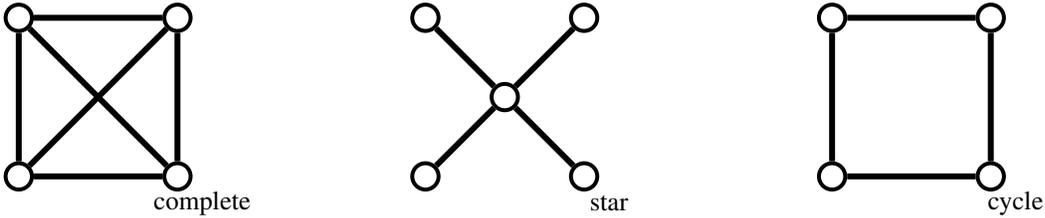

\begin{center}
\begin{minipage}[c]{.32\textwidth}
\includestandalone[width=0.5\linewidth]{./fig/regular-tikz} 
\end{minipage}
\begin{minipage}[c]{.32\textwidth}
\includestandalone[width=0.5\linewidth]{./fig/star-tikz} 
\end{minipage}
\begin{minipage}[c]{.32\textwidth}
\includestandalone[width=0.5\linewidth]{./fig/cycle-tikz} 
\end{minipage}
\begin{minipage}[c]{.32\textwidth}
\begin{center}
complete
\end{center}
\end{minipage}
\begin{minipage}[c]{.32\textwidth}
\begin{center}
star
\end{center}
\end{minipage}
\begin{minipage}[c]{.32\textwidth}
\begin{center}
cycle
\end{center}
\end{minipage}
\end{center}
\caption{A few typical network topologies.}
\label{fig:canonical-network}
\end{figure}

\cite{Sarkar.Roozbehani.ea:18} connects the macroscopic tail risk in $x_{\infty}$ to the (identity) \emph{Gramian} of $A$, while generalizing $A$ beyond $A=(1-\alpha)\mc W$. The Gramian matrix of $A$, denoted as $P(A)$, is the one that satisfies $P=A^T P A + I$. 
%
%
It is shown in \cite{Sarkar.Roozbehani.ea:18} that if each of $\{\omega_i\}_{i \in [n]}$ is exponentially-tailed with continuous, symmetric probability density function with full support, in addition to satisfying $\E[\omega]=\zerobf$ and $\E[\omega \omega^T]=I$, and if the network sequence $\{A_n\}_{n \in \naturals}$ is such that, for each $n$, (i) $A_n$ is non-negative, (ii) $A_n$ has a Perron root $\lambda_{\text{PF}}$, with corresponding right eigenvector $z$ satisfying $\max_{i} z_i/\min z_i = O(1)$, and (iii) a scaled version of the associated Gramian satisfies $\|P(A_n/\sqrt{\lambda_{\text{PF}}})\|_1=\Theta(1)$, then 
the network sequence does not have macroscopic tail risks. Accordingly, it can be shown that complete and cycle networks have no macroscopic tail risks, whereas star networks do; see Figure~\ref{fig:canonical-network} for an illustration of these network topologies. It is noted in \cite{Sarkar.Roozbehani.ea:18} that, conditions (i) and (ii) above on the networks allow $A_n=\gamma \mc W_n$ with $\gamma \in (0,1)$ and $\mc W_n$ a row-stochastic matrix, as well as other $A_n$ with $\|A_n\|_1<1$, and stable $A_n$ with $\|A_n\|_1>1$. This is a generalization of the setup in \cite{Acemoglu.Ozdaglar.ea:17}, where $A_n$ is restricted to a specific class of $A_n$ with $\|A_n\|<1$, satisfying $A_n = (1-\alpha) \mc W_n$, $\alpha \in (0,1)$. 

The Gramian also plays a role in characterizing networks robustness metrics related to the energy of the network, $\sum_{t=0}^{\infty} x^T(t) x(t)$. 
%
It is of interest to study the behavior of this quantity under deterministic and stochastic shocks $\omega$. Formally, the following quantities are of interest:
$$
\energymax(A) = \sup_{\|\omega\|_2=1} \, \sum_{t=0}^{\infty} x^T(t) x(t) \, , \qquad \energyavg(A) = \frac{1}{n} \mathbb{E}_{\omega}[\sum_{t=0}^{\infty} x^T(t) x(t)]
$$
where, in the definition of $\energyavg(A)$, $\mathbb{E}[\omega]=0$ and $\mathbb{E}[\omega \omega^T]=I$. $\energymax(A)$ and $\energyavg(A)$ are, respectively, referred to as the maximum and average disruption energy of $A$. These two quantities represent two different aspects of network robustness. Under a deterministic shock incident to the network, $\energymax(A)$ denotes the maximum energy that can propagate through the network, while, under a random shock, $\energyavg(A)$ denotes expected energy that propagates through the network. 

\ksmarginsoft{use $\mathbb{P}$ for Gramian to avoid confusion}
The Gramian gives a complete energy profile for a network. It is shown in \cite{Sarkar.Roozbehani.ea:18} that, for any $A$, $\energymax(A) = \sigma (P(A))$ and $\energyavg(A) = \frac{1}{n} \text{trace}(P(A))$. The latter is also related to the notion of $\mc H_2$-norm of a network, which measures the cumulative amplification of a shock due to network effects. Specifically, $\mc H_2(A)=\text{trace}(P(A))$. For any finite network, the two disruption energies are finite if $\rho(A)<1$. Therefore, a meaningful approach to characterize robustness is to consider a sequence $\{A_n\}_{n \in \naturals}$ of networks with $\rho(A_n)<1$ for every $n$, and study the scaling of disruption energies as the size $n \to \infty$.

It is shown in \cite{Sarkar.Roozbehani.ea:18} that, for large \emph{undirected} networks $A_n$, the disruption energies are related to the spectral radius as: $\energymax(A_n) = \frac{1}{1-\rho^2(A_n)}$ and $\energyavg(A_n) = O\left(\frac{1}{1-\rho(A_n)} \right)$. Specifically, for complete, star and cycle graphs (cf. Figure~\ref{fig:canonical-network}), $\energymax(A_n)$ scales as $\Theta(n)$, $\Theta(1)$ and $\Theta(1)$ respectively, and $\energyavg(A_n)$ scales as $\Theta(1)$, $\Theta(1)$ and $\Theta(n)$ respectively. On the other hand, for directed networks, the scaling of disruption energies  may be independent of spectra. In particular, there exist  vehicular platoon networks, e.g., see \cite{Herman.Martinec.ea:14}, for which $\mc H_2(A_n)$ is $\Omega(\exp(\delta n))$ for some $\delta>0$, even if $\rho(A_n)$ is uniformly bounded away from $1$. These results suggest possible advantage of undirected networks over directed networks. Indeed, it is shown in \cite{Sarkar.Roozbehani.ea:18} that making a large directed network $A_n$ more undirected does not increase the two disruption energies in an order sense, and provides example where it decreases them. Such a process of making a network more undirected is achieved by \emph{spectral balancing} of its $A_n$. Formally, for $\epsilon \in (0,1)$, $\epsilon-$balanced version of $A_n$ is $A_n^{\varepsilon}=(1-\varepsilon) A_n + \varepsilon U_n \Gamma_n U_n^T$, where $\Gamma_n$ is a diagonal matrix such that $\rho(\Gamma_n) \leq \rho(A_n)$ and $U_n$ is obtained from the spectral decomposition of the Gramian of $A_n$: $P(A_n)=U_n D_n U_n^T$.


\paragraph{Future Research Directions}
The dependence of macroscopic tail risks and disruption energies on Gramian suggests a connection between these two notions from network robustness perspective. However, the exact relationship remains to be established.

%% file: fig/star-tikz.tex
\begin{tikzpicture}

\path (-1,1) node(a) [circle,draw,line width=0.5mm] {$  $}
	(1,1) node(b) [circle,draw,line width=0.5mm] {$  $}
	(1,-1) node(c) [circle,draw,line width=0.5mm] {$  $}
	(-1,-1) node(d) [circle,draw,line width=0.5mm] {$  $}
	(0,0) node(e) [circle,draw,line width=0.5mm] {$  $};

\draw[very thick, line width=0.75mm] (a) -- (e) -- (c);
\draw[very thick, line width=0.75mm] (b) -- (e) -- (d);
\end{tikzpicture}

%% file: fig/cycle-tikz.tex
\begin{tikzpicture}

\path (-1,1) node(a) [circle,draw,line width=0.5mm] {$  $}
	(1,1) node(b) [circle,draw,line width=0.5mm] {$  $}
	(1,-1) node(c) [circle,draw,line width=0.5mm] {$  $}
	(-1,-1) node(d) [circle,draw,line width=0.5mm] {$  $};

\draw[very thick, line width=0.75mm] (a) -- (b) -- (c) -- (d) -- (a);

\end{tikzpicture}

%% file: conclusion.tex
\section{Conclusion}
We presented an overview of multiple approaches to analyzing robustness of networked systems. The common theme is to highlight the role played by the underlying network structure in determining robustness. 

For linear dynamics over networks with unreliable links, the interplay of the reliability of links and their role in the dynamics, as captured by the \emph{loop gain operator} of the associated covariance feedback system, determines whether the network dynamics could or could not arbitrarily amplify additive white noise. In economic network setting, heterogeneity in the node degrees plays a key role in determining the rate at which the deviation in the aggregate equilibrium network output, due to independent shocks to individual nodes, decays to zero as the network size increases. On the other hand, the interplay between the distribution functions of the individual shocks and the heterogeneity in node dominance determines whether the aggregate output exhibits tail risk. The aggregate output in economic network at equilibrium can be related to the transient of a related linear network dynamics subject to initial shock. Such an abstraction allows to relate macroscopic tail risks to the corresponding Gramian, and allows to study tail risk beyond economic network settings. The Gramian also plays a key role in determining scaling of network disruption energies which serve as meaningful robustness metrics.  

For physical flow over networks, the notions of link, node and network residual capacities are key determinants of robustness to reduction in capacity. These quantities are relatively fast to compute in the static setting when the flow is physically constrained only by Kirchoff law. In presence of additional constraint, such as Ohm's law in electrical networks, computing the relevant quantities is hard in general; however, tree reducible structure of the network helps to considerably reduce the complexity. In the dynamical setting, network robustness is known to be influenced by information constraint and nature of control actions. For instance, decentralized control does not cause loss in robustness in certain scheduling scenarios, but it does so in certain routing scenarios. In the latter case, adding scheduling to control action allows to recover the loss in robustness. 

The extension of network flow robustness to cascading failure setting involves dynamic programming like computations, with considerable simplifications possible under tree like structure of the network and symmetry about source-sink pair. For contagion-based cascading failure process, the maximum, among all nodes, likelihood of failure is determined by the interplay of the node threshold assignment and the network structure. Within regular graphs, the clique structure and the complete $d$-ary tree structure provide a clean way to analyze this interplay, if the node degrees are two or if the node threshold values are upper bounded by two. 

\kscommentjss{The selection of results and setups reviewed in the paper naturally reflect our bias stemming from our work in this area. Therefore, the material in the paper is to be interpreted as reflecting a subset of the growing literature on network robustness.}